\documentclass[11pt,eqs,twocolumn]{article}
\usepackage{latexsym,amsmath,amstext,amssymb,mathtools}
\usepackage{graphicx}
\usepackage[all]{xy}

\def\cleq{\setcounter{equation}{0}}

\usepackage[margin=0.5in]{geometry}

\title{
T-duality diagram for a weakly curved background
\thanks{Work supported in part by
the Serbian Ministry of Education, Science and Technological Development, under contract No. 171031.}}
\author{Lj. Davidovi\'c\thanks{e-mail: ljubica@ipb.ac.rs} ,
B. Nikoli\'{c}\thanks{e-mail: bnikolic@ipb.ac.rs}\,
and B. Sazdovi\'c\thanks{e-mail: sazdovic@ipb.ac.rs}\\
{\it Institute of Physics,}\\
{\it University of Belgrade,}\\
{\it 11001 Belgrade, P.O.Box 57, Serbia}
}

\begin{document}
\maketitle

\begin{abstract}
In one of our previous papers we generalized the Buscher T-dualization procedure.
Here we will investigate the application of this procedure to the
theory of a bosonic string moving in the weakly curved background.
We obtain the complete T-dualization diagram,
connecting the theories which are the result of the T-du\-a\-li\-za\-tions
over all possible choices of the coordinates.
We distinguish three forms of the T-dual theories: the initial theory, the theory
obtained T-dualizing some of the coordinates of the initial theory
and the theory obtained T-dualizing all of the initial coordinates.
While the initial theory is geometric, all the other theories are non-geometric and additionally non-local.
We find the T-dual coordinate transformation laws connecting these theories
and show that the set of all T-dualizations forms an Abelian group.

\end{abstract}

\section{Introduction}
\cleq

T-duality is a property of string theory that was not encountered in any point particle theory \cite{GPR,AAL,Z,BBSk}.
Its discovery was surprising,
because it implies that there exist theories, defined for essentially 
different geometries of the compactified dimensions,
which are physically equivalent.
The origin of T-duality is seen
in the possibility that,
unlike a point particle the string can wrap around compactified dimensions. 
But, no matter if one dimension is compactified on a circle 
of radius $R$  or rather on a circle of radius $l_s^2/R$, where $l_s$ is the fundamental string length scale,
the theory will
describe the string with the same physical properties.
The investigation of T-duality does not cease to provide interesting new physical implications.

The prescription for obtaining the equivalent T-dual theories is given by the Buscher T-dualization procedure \cite{B,RV}.
The procedure is applicable along the isometry directions,
which allows the investigation of the backgrounds which do not depend on some coordinates.
It is found that T-duality transforms geometric backgrounds to the non-geometric backgrounds with $Q$ flux
which are locally well defined,
and these to different types of non-geometric backgrounds, backgrounds with $R$ flux which are not well defined even locally \cite{TDC,BDLPR}. A similar prescription can be used to obtain fermionic T-duality \cite{BBS}.
It is argued that the better understanding of T-duality should be sought for by
doubling the coordinates, investigating the theories in which the background fields depend on both the usual space-time coordinates
and their doubles \cite{H,HZ,HHZ,HHZ1}, which would make the T-duality a manifest symmetry.

T-duality enables the investigation of the closed st\-r\-ing non-commutativity.
The coordinates of the closed string are commutative when the string moves in a constant background.
In a 3-dimensional space with
the Kalb-Ramond field depending on one of the coordinates, successive T-dualizations along isometry directions lead to a theory with 
Q flux and the non-commutative coordinates \cite{L,ALLP,ALLP2}. 
The novelty in the research is the generalized T-dualization procedure,
realized in \cite{DS}, addressing  the bosonic string moving in the weakly curved background - constant gravitational field and coordinate dependent Kalb-Ra\-m\-ond field with an infinitesimal field strength.
The non-commutativity characteristics of a closed string moving in the weakly curved background was
considered in \cite{DNS}.

The generalized procedure is applicable to all the space-time coordinates on which the string backgrounds depend.
In Ref. \cite {DS}, it was
first applied to all initial coordinates, which produces a T-dual theory;
it was then applied to all the T-dual coordinates and the initial theory was obtained.
In this paper,
we will investigate the application of the generalized T-du\-ali\-za\-tion procedure
to an arbitrary set of coordinates.
Let us denote
the T-dualization along direction $x^\mu$
 by $T^\mu$ and the T-dualization along dual direction $y_\mu$ by  $T_\mu$.
Choosing $d$ arbitrary directions, we denote
\begin{eqnarray}\label{eq:ta}
{\cal T}^{a}=\circ_{n=1}^{d}T^{\mu_{n}},\quad&
{\cal T}^{i}=\circ_{n=d+1}^{D}T^{\mu_{n}},\quad&
{\cal T}=\circ_{n=1}^{D}T^{\mu_{n}},
\nonumber\\
\end{eqnarray}
\begin{eqnarray}\label{eq:tainv}
{\cal T}_{a}=\circ_{n=1}^{d}T_{\mu_{n}},\quad&
{\cal T}_{i}=\circ_{n=d+1}^{D}T_{\mu_{n}},\quad&
\widetilde{\cal T}=\circ_{n=1}^{D}T_{\mu_{n}},
\nonumber\\
\end{eqnarray}
where $\mu_{n}\in (0,1,\dots,D-1)$,
and $\circ$ denotes the composition of T-dualizations.
We will apply T-dualizations (\ref{eq:ta}) to
the initial theory, and T-dualizations (\ref{eq:tainv})
to its completely T-dual theory (obtained in \cite{DS}). We will prove the 
following  composition laws:
\begin{eqnarray}
{\cal T}^i\circ{\cal T}^a={\cal T},
\quad
{\cal T}_i\circ{\cal T}_a=\widetilde{\cal T},
\quad
{\cal T}_a\circ{\cal T}^a=1,
\end{eqnarray}
where $1$ denotes the identical transformation (T-du\-ali\-za\-tion not performed).
So, elements $1,{\cal T}^{a}$ and ${\cal T}_{a}$, with $d=1,\dots,D$,
form an Abelian group.
We will find the explicit form of the resulting theories and the
corresponding T-dual coordinate transformation laws.
These results complete the T-dualization diagram
connecting all the theories T-dual to the initial theory.

Throughout the whole article (except for Sect. 9)  we assume that the Kalb-Ramond field depends on all coordinates. In that case all T-dual theories, except the initial theory, are non-geometric and  non-local because they depend on variable $V^\mu$,
which is a line integral of the derivatives of the dual coordinates.
To all of these theories there corresponds a flux which is of the same type as the $R$ flux unlike the 
non-geometric theories with $Q$ flux,
which have a local geometric description.

In Sects. 9.1 and 9.2,  we present an example of the
$3$-dimensional torus, $T^3$ with H-flux, where Kalb-Ramond field depends only on coordinate $x^3$. Then T-dualizations along isometry directions $x^1$ and $x^2$ lead to geometric background and the T-dualization along  
$x^3$ leads to non-geometric background.
In Sect. 9.1 putting $D=3$, $d=1,2$ with $B_{\mu\nu}$ depending on $x^3$ we reproduce the T-duality chain of Refs. \cite{L,ALLP,ALLP2}.

In Sect. 9.2 we will compare the results of our paper with those of Ref. \cite{BDLPR}.
In our manuscript,
the background fields' argument, the variable $V^\mu$,
incorporates all features of the non-geometric spaces. First,  as pointed out in Ref. \cite{BDLPR} it "eludes a geometric description even locally"  because it is a line integral of the derivative. 
Second, we obtain non-associativity 
and breaking of Jacobi identity 
typical for the so called R-flux backgrounds. In Sect. 9.3 we present example of the 4-dimensional torus $T^4$ to generalize the case of Ref. \cite{GRV} to critical surface.

The generalized T-dualization procedure
originates from the Buscher T-dualization procedure.
The first rule in the prescription is to replace
the de\-ri\-va\-ti\-ves with the covariant derivatives.
The new point in the prescription is the replacement
of the coordinates in the background fields' argument
with the invariant coordinates.
The invariant coordinates are 
defined as the line integrals of the covariant derivatives of the original coordinates.
Both covariant derivatives and invariant coordinates are defined using the gauge fields.
These fields should be nonphysical,
so one requires that their field strength should be zero.
This is realized by adding the 
corresponding Lagrange multipliers' terms. As a consequence of the translational symmetry one can fix the coordinates along which the T-dualization is performed and obtain a gauge fixed action.
An important cross-way in the T-dualization procedure
is determined by the equations of motion of the gauge fixed action.
Two equations of motion obtained varying this action are used to direct the procedure
either back to the initial action or forward to the T-dual action.
For the equation of motion obtained varying the action over the Lagrange multipliers,
the gauge fixed action reduces to the initial action.
For the equation of motion obtained varying the action over the gauge fields one obtains
the T-dual theory.
Comparing the solutions for the gauge fields in these two directions,
one obtains the T-dual coordinate transformation laws.


\section{T-duality in the weakly curved background}
\cleq
Let us consider the closed bosonic string propagating in the background with metric field $G_{\mu\nu}$,
Kalb-Ramond field
$B_{\mu\nu}$ and a dilaton field $\Phi$, described by the action \cite{Z,BBSk}
\begin{eqnarray}\label{eq:action0}
S[x]&=& \kappa \int_{\Sigma} d^2\xi\sqrt{-g} \Big[\Big(\frac{1}{2}{g}^{\alpha\beta}G_{\mu\nu}(x)
+\frac{\varepsilon^{\alpha\beta}}{\sqrt{-g}}B_{\mu\nu}(x)\Big)\!\cdot
\nonumber\\
&&\cdot
\partial_{\alpha}x^{\mu}\partial_{\beta}x^{\nu}+\frac{1}{4\pi\kappa}\Phi(x)R^{(2)}\Big].
\end{eqnarray}
The integration goes over a 2-dimensional world-sheet $\Sigma$
parametrized by
$\xi^\alpha$ ($\xi^{0}=\tau,\ \xi^{1}=\sigma$),
$g_{\alpha\beta}$ is the intrinsic world-sheet metric, $R^{(2)}$ corresponding 2-di\-me\-nsi\-o\-nal scalar curvature,
$x^{\mu}(\xi),\ \mu=0,1,...,D-1$ are the coordinates of the
D-dimensional space-time,
$\kappa=\frac{1}{2\pi\alpha^\prime}$
with $\alpha^\prime$ being the Regge slope parameter
and $\varepsilon^{01}=-1$.

\subsection{Weakly curved background}

The requirement of the quantum conformal
invariance of the world-sheet results
in  the space-time equations of motion for the
background fields.
In the lowest order in the slope parameter $\alpha^\prime$ these equations are
\begin{eqnarray}\label{eq:beta}
&&
R_{\mu\nu}-\frac{1}{4}B_{\mu\rho\sigma}B_\nu^{\ \rho\sigma}+2D_\mu \partial_\nu\Phi=0,
\nonumber\\
&&
D_\rho B^\rho_{\ \mu\nu}-2\partial_\rho\Phi B^\rho_{\ \mu\nu}=0,
\nonumber\\
&&
4(\partial\Phi)^{2}-4D_\mu\partial^\mu\Phi
+\frac{1}{12}B_{\mu\nu\rho}B^{\mu\nu\rho}
+4\pi\kappa(D-26)/3
\nonumber\\
&&
-R
=0\,.
\end{eqnarray}
Here
$B_{\mu\nu\rho}=\partial_\mu B_{\nu\rho}
+\partial_\nu B_{\rho\mu}+\partial_\rho B_{\mu\nu}$
is the field strength of the field $B_{\mu \nu}$, and
$R_{\mu \nu}$ and $D_\mu$ are the Ricci tensor and the
covariant derivative with respect to the space-time metric.
We will consider
one of the simplest coordinate dependent solutions
of (\ref{eq:beta}),
the weakly curved background.
This background was considered in Refs. \cite{DS1,DS2,DS3},
where the influence of the boundary conditions on the non-commutativity
of the open bosonic string has been investigated.
The same approximation
was considered in \cite{ALLP,DNS} in context of the closed string non-commutativity.

The weakly curved background is defined by
\begin{eqnarray}\label{eq:wcb}
&&G_{\mu\nu}(x)=const,
\nonumber\\
&&B_{\mu\nu}(x)=b_{\mu\nu}+
\frac{1}{3}B_{\mu\nu\rho}x^\rho
\equiv 
b_{\mu\nu}+
h_{\mu\nu}(x),
\nonumber\\
&&\Phi(x)=const,
\end{eqnarray}
with $b_{\mu\nu},B_{\mu\nu\rho}=const$.
This background is the solution of 
the space-time equations of motion
if the constant
$B_{\mu\nu\rho}$ is taken to be infinitesimal
and all the calculations are done in the first order
in $B_{\mu\nu\rho}$,
so that the curvature $R_{\mu\nu}$ can be neglected as the infinitesimal of the second order.
Through the whole manuscript (with the exeption of Sect. 9) we assume that the background has topology of $D$-dimensional torus $T^D$, where Kalb-Ramond field depends
on all coordinates. In Sects. 9.1 and 9.2  we give an example of the $3$-dimensional torus, $T^3$, with H-flux, where the Kalb-Ramond field depends only on  coordinate $x^3$, while in Sect. 9.3 we give an example of 4-dimensional torus $T^4$ with constant background fields.

The assumption that
$B_{\mu\nu\rho}$ is infinitesimal means that we consider the $D$-dimensional torus so large that for any choice of indices holds \cite{ALLP}
\begin{equation}
\frac{B_{\mu\nu\rho}}{R_\mu R_\nu R_\rho}\ll 1\, ,
\end{equation}
where $R_\mu$ are the radii of the torus. The $H$-flux background, considered in Refs. \cite{BDLPR,ALLP}, is of the same type as the weakly curved background. However, this background 
depends just on $x^3$ and corresponds to the examples addressed in Sect. 9 of our paper. The background considered in the rest of the article depends on all coordinates.

In this paper we will investigate the T-dualization properties of the action (\ref{eq:action0})
describing the closed string moving in the weakly curved background.
Taking the conformal gauge $g_{\alpha\beta}=e^{2F}\eta_{\alpha\beta}$, the action (\ref{eq:action0}) becomes
\begin{equation}\label{eq:action1}
S[x] = \kappa \int_{\Sigma} d^2\xi\
\partial_{+}x^{\mu}
\Pi_{+\mu\nu}(x)
\partial_{-}x^{\nu},
\end{equation}
with the background field composition equal to
\begin{equation}\label{eq:pi}
\Pi_{\pm\mu\nu}(x)=
B_{\mu\nu}(x)\pm\frac{1}{2}G_{\mu\nu}(x),
\end{equation}
and the light-cone coordinates given by
\begin{equation}
\xi^{\pm}=\frac{1}{2}(\tau\pm\sigma),
\qquad
\partial_{\pm}=
\partial_{\tau}\pm\partial_{\sigma}.
\end{equation}

\subsection{Complete T-dualization}

The T-dualization of the
closed string theory in the weakly curved background
was presented in \cite{DS}.
The procedure is related to a global symmetry of the theory
\begin{equation}\label{eq:globalna}
\delta x^\mu=\lambda^\mu.
\end{equation}
The symmetry still exists in the presence of the nontrivial Kalb-Ramond field (\ref{eq:wcb}),
but only in the case of the trivial mapping of the world-sheet into the space-time,
because in that case
the variation of the action (\ref{eq:action1})
\begin{equation}
\delta S=\frac{\kappa}{3}\varepsilon^{\alpha\beta} B_{\mu\nu\rho} 
\lambda^\rho\int d^2\xi  \partial_\alpha x^\mu \partial_\beta x^\nu
\end{equation}
after partial integration, using identity $\varepsilon^{\alpha\beta}\partial_\alpha\partial_\beta=0$, becomes
\begin{equation}\label{eq:vars}
\delta S=
\frac{\kappa}{3}B_{\mu\nu\rho}\lambda^\rho
\epsilon^{\alpha\beta}
\int d^{2}\xi\partial_{\alpha}(x^\mu\partial_{\beta}x^\nu),
\end{equation}
which is equal to zero.
This means that classically, directions which appear in the argument of Kalb-Ramond field are also Killing directions.
However the standard Buscher procedure cannot be applied to them,
because background fields depend on the coordinates but not on their derivatives.

The T-dual picture of the theory,
obtained applying the T-dualization procedure to all the coordinates,
is given by
\begin{eqnarray}\label{eq:dualna}
S[y]&=&
\kappa
\int d^{2}\xi\
\partial_{+}y_\mu
\,^\star \Pi_{+}^{\mu\nu}\big(\Delta V(y)\big)\,
\partial_{-}y_\nu
\nonumber\\
&=&
\frac{\kappa^{2}}{2}
\int d^{2}\xi\
\partial_{+}y_\mu
\Theta_{-}^{\mu\nu}\big(\Delta V(y)\big)
\partial_{-}y_\nu,
\end{eqnarray}
where
\begin{eqnarray}\label{eq:tetapm}
{\Theta}^{\mu\nu}_{\pm}&\equiv&
-\frac{2}{\kappa}
(G^{-1}_{E}\Pi_{\pm}G^{-1})^{\mu\nu}=
{\theta}^{\mu\nu}\mp \frac{1}{\kappa}(G_{E}^{-1})^{\mu\nu}\!,
\end{eqnarray}
with
\begin{eqnarray}\label{eq:effpolja}
G_{E\mu\nu}&\equiv& G_{\mu\nu}-4(BG^{-1}B)_{\mu\nu},
\nonumber\\
\theta^{\mu\nu}&\equiv&
-\frac{2}{\kappa}
(G^{-1}_{E}BG^{-1})^{\mu\nu},
\end{eqnarray}
being the effective metric and the non-commutativity parameter in Seiberg-Witten terminology of the open bosonic string theory \cite{SW}.
The T-dual background fields  are equal to
\begin{eqnarray}\label{eq:dpolja}
&&^\star G^{\mu\nu}\big(\Delta V(y)\big)=
(G_{E}^{-1})^{\mu\nu}\big(\Delta V(y)\big),
\nonumber\\
&&
^\star B^{\mu\nu}\big(\Delta V(y)\big)=
\frac{\kappa}{2}
{\theta}^{\mu\nu}\big(\Delta  V(y)\big),
\end{eqnarray}
and their argument is given by
\begin{eqnarray}\label{eq:vargdef}
\Delta V^\mu(y)&=&-\frac{\kappa}{2}\left(\Theta^{\mu\nu}_{0-}
+\Theta^{\mu\nu}_{0+}\right)\Delta y_\nu
\nonumber\\
&&
+\frac{\kappa}{2}\left(\Theta_{0-}^{\mu\nu}-\Theta^{\mu\nu}_{0+}\right)\Delta{\tilde{y}}_\nu
\nonumber \\
&=&-\kappa\theta_{0}^{\mu\nu}\Delta y_\nu+(g^{-1})^{\mu\nu}\Delta{\tilde{y}}_\nu.
\end{eqnarray}
Here
$\Theta_{0\pm}^{\mu\nu}$ is the zeroth order value of the field composition $\Theta_{\pm}^{\mu\nu}$ defined in (\ref{eq:tetapm})
and $g_{\mu\nu}=G_{\mu\nu}-4b^{2}_{\mu\nu}$
and ${\theta}^{\mu\nu}_{0}
=-\frac{2}{\kappa}
(g^{-1}bG^{-1})^{\mu\nu}
$ are the zeroth order values of the effective fields (\ref{eq:effpolja}).
The variable $\Delta{\tilde{y}}_\mu$
is the double of the dual variable $\Delta y_\mu=y_\mu(\xi)-y_\mu(\xi_{0})$,
defined as the following line integral:
\begin{equation}\label{eq:dualnavar}
\Delta{\tilde{y}}_\mu=\int_{P}(d\tau y^\prime_\mu+d\sigma \dot{y}_\mu)
=\int_{P} d\xi^\alpha\varepsilon^\beta_{\ \alpha}\partial_\beta y_\mu,
\end{equation}
taken
along the path $P$,
from the point $\xi^{\alpha}_{0}(\tau_{0},\sigma_{0})$
to the point $\xi^{\alpha}(\tau,\sigma)$.

The fact that we are working with the weakly curved background ensures that the T-dual background fields are the solution of the space-time equations
(5). Because both dual metric ${}^\star G^{\mu\nu}$ and dual Kalb-Ramond field ${}^\star B^{\mu \nu}$ are linear in coordinates with infinitesimal coefficients, the
dual Christoffel symbol  ${}^\star \Gamma_\mu^{\nu \rho}$ and dual field strength  ${}^\star B^{\mu \nu \rho}$ are constant and infinitesimal.
In Eq. (\ref{eq:diltrans}) of Sect. 8 we will show that T-dual dilaton field
is ${}^\bullet \Phi=\Phi-\ln\det\sqrt{2\Pi_{+}}$, where $\Phi$ is constant and $\Pi_{+}$ is linear in coordinates with infinitesimal coefficients. So, ${}^\bullet \Phi$ is also linear in coordinates with infinitesimal coefficients, and
$\partial_\mu {}^\bullet\Phi$ is constant and infinitesimal. Consequently, $D_\mu\partial_\nu {}^\bullet\Phi$, $\partial_\rho {}^\bullet\Phi B^\rho{}_{\mu\nu}$ and $(\partial_\mu {}^\bullet\Phi)^2$ 
are infinitesimals 
of the second order. So, all T-dual space-time equations, for the metric, for the Kalb-Ramond field and for dilaton field, are infinitesimals of the second order and as such are neglected.

The initial theory (\ref{eq:action1}) and its completely T-dual theory (\ref{eq:dualna})
 are connected by the T-dual coordinate transformation laws (eq. (42) of Ref. \cite{DS})
\begin{equation}\label{eq:zak}
\partial_\pm x^\mu=-\kappa \Theta^{\mu\nu}\big(\Delta V\big)\partial_\pm y_\nu\mp2\kappa\Theta^{\mu\nu}_{0\pm}\beta^\mp_\nu\big(V\big),
\end{equation}
and its inverse (eq. (66) of Ref. \cite{DS})
\begin{equation}\label{eq:zakinv}
\partial_{\pm}y_\mu\cong
-2\Pi_{\mp\mu\nu}(\Delta x)\partial_{\pm}x^\nu\mp
2\beta^{\mp}_{\mu}(x),
\end{equation}
where $\beta^{\pm}_\mu(x)=
\mp\frac{1}{2}h_{\mu\nu}(x)\partial_\mp x^\nu$.
It is shown that
\begin{equation}
{\cal T}: S[x^\mu]\to S[y_\mu],\quad
\widetilde{\cal T}: S[y_\mu]\to S[x^\mu],
\end{equation}
and therefore
\begin{equation}
\mathcal T\circ \widetilde{\mathcal T}=1.
\end{equation}

\section{T-dualization along arbitrary subset of coordinates
 ${\cal T}^{a}:S[x^\mu]\rightarrow S[x^{i},y_{a}]$}\label{sec:prvi}
 \cleq

In this section, we will learn what theory is obtained
if one chooses to  apply the T-dualization procedure
to the action (\ref{eq:action1}),
along arbitrary $d$ coordinates $x^{a}$, ${\cal T}^{a}:S[x^\mu]\rightarrow S[x^{i},y_{a}]$,
with ${\cal T}^{a}=\circ_{n=1}^{d}T^{\mu_{n}}$, 
$\mu_{n}\in (0,1,\dots,D-1)$.

The closed string action in the weakly curved background (\ref{eq:wcb}) has a global
symmetry (\ref{eq:globalna}).
One localizes the symmetry
for the coordinates $x^{a}$,
by introducing the
gauge fields $v^{a}_{\alpha}$ and
substituting the ordinary derivatives with the covariant derivatives
\begin{equation}
\partial_{\alpha}x^{a}\rightarrow
D_{\alpha}x^{a}=\partial_\alpha x^{a}
+v^{a}_{\alpha}.
\end{equation}
The covariant derivatives are invariant under standard gauge transformations
\begin{equation}\label{eq:gft}
\delta v^{a}_{\alpha}=-\partial_\alpha \lambda^{a}.
\end{equation}
In the case of the weakly curved background,
in order to obtain the gauge invariant action
one should additionally substitute the coordinates $x^{a}$ in the argument of the background fields with their invariant extension,
defined by
\begin{eqnarray}
\Delta x^{a}_{inv}&\equiv&
\int_{P}d\xi^\alpha\,D_\alpha x^{a}
=\int_{P}(d\xi^{+}D_{+}x^{a}
+d\xi^{-}D_{-}x^{a})
\nonumber\\&=&
x^{a}-x^{a}(\xi_{0})
+\Delta V^{a},
\end{eqnarray}
where
\begin{equation}\label{eq:vdef}
\Delta V^{a}
\equiv
\int_{P}d\xi^\alpha v^{a}_{\alpha}
=\int_{P}(d\xi^{+} v^{a}_{+}
+d\xi^{-} v^{a}_{-}).
\end{equation}
To preserve the physical equivalence
between the gauged and the original theory,
one introduces the Lagrange multipliers $y_{a}$ and adds term
$\frac{1}{2}y_{a} F^{a}_{+-}$
 to
the Lagrangian,
which will force the field strength $F^{a}_{+-}\equiv
\partial_{+}v^{a}_{-}
-\partial_{-}v^{a}_{+}=-2F^{a}_{01}$ to vanish.
In this way, the gauge invariant action
\begin{eqnarray}\label{eq:ainv}
&&S_{inv}[x^\mu,x^{a}_{inv},y_{a}]=
\nonumber\\
&&\kappa\int d^{2}\xi\Big{[}
\partial_{+}x^{i}\Pi_{+ij}\big(x^{i},\Delta x^{a}_{inv}\big)\partial_{-}x^{j}
\nonumber\\
&&\quad+\partial_{+}x^{i}\Pi_{+ia}\big(x^{i},\Delta x^{a}_{inv}\big)D_{-}x^{a}
\nonumber\\
&&\quad+D_{+}x^{a}\Pi_{+ai}\big(x^{i},\Delta x^{a}_{inv}\big)\partial_{-}x^{i}
\nonumber\\
&&\quad+D_{+}x^{a}\Pi_{+ab}\big(x^{i},\Delta x^{a}_{inv}\big)D_{-}x^{b}
\nonumber\\
&&\quad+\frac{1}{2}(v^{a}_{+}\partial_{-}y_{a}
-v^{a}_{-}\partial_{+}y_{a})
\Big{]}
\end{eqnarray}
is obtained,
where the last term is equal to
$\frac{1}{2}y_{a} F^{a}_{+-}$ up to the total divergence.
Now, we can fix the gauge taking
$x^{a}(\xi)=x^{a}(\xi_{0})$ and obtain the gauge fixed action
\begin{eqnarray}\label{eq:sfix}
&&S_{fix}[x^{i},v^{a}_\pm,y_{a}]=
\nonumber\\
&&
\kappa\int d^{2}\xi\Big{[}
\partial_{+}x^{i}\Pi_{+ij}\big(x^{i},\Delta V^{a}\big)\partial_{-}x^{j}
\nonumber\\
&&\quad
+\partial_{+}x^{i}\Pi_{+ia}\big(x^{i},\Delta V^{a}\big)v_{-}^{a}
+v_{+}^{a}\Pi_{+ai}\big(x^{i},\Delta V^{a}\big)\partial_{-}x^{i}
\nonumber\\
&&\quad
+v_{+}^{a}\Pi_{+ab}\big(x^{i},\Delta V^{a}\big)v_{-}^{b}
+\frac{1}{2}(v^{a}_{+}\partial_{-}y_{a}
-v^{a}_{-}\partial_{+}y_{a})
\Big{]}.
\nonumber\\
\end{eqnarray}
This action reduces to the initial one for the equations of motion obtained varying over the Lagrange multipliers.
The T-dual action is obtained for the equations of motion for the gauge fields.

\subsection{Regaining the initial action}

Varying the gauge fixed action (\ref{eq:sfix}) over
the Lagrange multipliers $y_{a}$ one obtains
the equations of motion
\begin{equation}\label{eq:emy}
\partial_{+}v^{a}_{-}
-\partial_{-}v^{a}_{+}=0,
\end{equation}
which have the solution
\begin{equation}\label{eq:vsola}
v^{a}_{\pm}=\partial_{\pm}x^{a}.
\end{equation}
For this solution the background fields' argument $\Delta V^{a}$ defined in (\ref{eq:vdef})
is path independent and reduces to
\begin{equation}\label{eq:vvelx}
\Delta V^{a}(\xi)=x^{a}(\xi)-x^{a}(\xi_{0}).
\end{equation}
The gauge fixed action (\ref{eq:sfix}) reduces to the initial action (\ref{eq:action1}),
but the background fields' argument is $\Delta V^{a}$ instead of $x^{i}$.
However, the action (\ref{eq:action1}) is invariant under the constant shift of coordinates,
so shifting coordinates by $x^{a}(\xi_{0})$ one obtains the exact form of the initial action.

\subsection{The T-dual action}\label{sec:pscrtom}

Using the equations of motion for the gauge fields, we eliminate them and obtain the T-dual action.

The equations of motion obtained varying 
the gauge fixed action (\ref{eq:sfix}) 
over the gauge fields $v^{a}_{\pm}$ are
\begin{eqnarray}\label{eq:emgf}
&&\Pi_{\pm ai}\big(x^{i},\Delta V^{a}\big)\partial_{\mp}x^{i}
+\Pi_{\pm ab}\big(x^{i},\Delta V^{a}\big)v^{b}_{\mp}
+\frac{1}{2}\partial_\mp y_{a}=
\nonumber\\
&&
\pm\beta^{\pm}_{a}\big(x^{i},V^{a}\big),
\end{eqnarray}
where
\begin{eqnarray}\label{eq:betaprva}
\beta^{\pm}_{a}\big(x^{i},V^{a}\big)&=&
\mp\frac{1}{2}
\Big{[}
h_{ai}(x^{i})\partial_{\mp}x^{i}
+h_{ab}(x^{i})\partial_{\mp}V^{b}
\nonumber\\
&+&h_{ai}\big(V^{a}\big)\partial_{\mp}x^{i}
+h_{ab}\big(V^{a}\big)\partial_{\mp}V^{b}
\Big{]}
\end{eqnarray}
is the contribution from the background fields' argument $\Delta V^{a}$,
defined in a same way as in Ref. \cite{DS},
by 
$\delta_{V}S_{fix}=-\kappa\int d^{2}\xi(\beta^{+}_{a}\delta v^{a}_{+}+\beta^{-}_{a}\delta v^{a}_{-})$.
If the initial background $\Pi_{+\mu\nu}$ does not depend on the coordinates $x^{a}$, the corresponding beta functions are zero
$\beta^{\pm}_{a}=0.$

Multiplying Eq. (\ref{eq:emgf}) by $2\kappa\tilde{\Theta}^{ab}_{\mp}$, defined in (\ref{eq:barteta1}), the inverse of the background fields composition $\Pi_{\pm ab}$, one obtains
\begin{eqnarray}\label{eq:vres}
v^{a}_{\mp}&=&-2\kappa\tilde{\Theta}^{ab}_{\mp}
\big(x^{i},\Delta V^{a}\big)\Big{[}
\Pi_{\pm bi}\big(x^{i},\Delta V^{a}\big)\partial_{\mp}x^{i}
+\frac{1}{2}\partial_\mp y_{b}
\nonumber\\
&&
\mp\beta^{\pm}_{b}\big(x^{i},V^{a}\big)
\Big{]}.
\end{eqnarray}
Substituting (\ref{eq:vres}) into the action (\ref{eq:sfix}),
we obtain
the T-dual action
\begin{eqnarray}\label{eq:tdualdel}
&&S[x^{i},y_{a}]=
\nonumber\\
&&\kappa\int d^{2}\xi\bigg{[}
\partial_{+}x^{i}{\overline{\Pi}}_{+ij}\big(x^{i},\Delta V^{a}(x^{i},y_{a})\big)
\partial_{-}x^{j}
\nonumber\\
&&\quad-\kappa\,\partial_{+}x^{i}\Pi_{+ia}\big(x^{i},\Delta V^{a}(x^{i},y_{a})\big)\!\cdot
\nonumber\\
&&\qquad\cdot
\tilde\Theta^{ab}_{-}\big(x^{i},\Delta V^{a}(x^{i},y_{a})\big)\partial_{-}y_{b}
\nonumber\\
&&\quad+\kappa\,\partial_{+}y_{a}\tilde\Theta^{ab}_{-}\big(x^{i},\Delta V^{a}(x^{i},y_{a})\big)\!\cdot
\nonumber\\
&&\qquad\cdot
\Pi_{+bi}\big(x^{i},\Delta V^{a}(x^{i},y_{a})\big)\partial_{-}x^{i}
\nonumber\\
&&\quad+\frac{\kappa}{2}\,\partial_{+}y_{a}\tilde\Theta^{ab}_{-}\big(x^{i},\Delta V^{a}(x^{i},y_{a})\big)\partial_{-}y_{b}
\bigg{]},
\end{eqnarray} 
where
\begin{equation}\label{eq:barpi}
{\overline{\Pi}}_{+ij}\equiv
\Pi_{+ij}
-2\kappa\Pi_{+ia}\tilde\Theta^{ab}_{-}\Pi_{+bj}.
\end{equation}

In order to find the explicit value of the background fields argument $\Delta V^{a}(x^{i},y_{a})$,
it is enough to consider the zeroth order of the equations of 
motion for the gauge fields $v^{a}_{\pm}$ (\ref{eq:vres})
\begin{eqnarray}\label{eq:vnulares}
v^{(0)a}_{\pm}=-2\kappa\tilde\Theta^{ab}_{0\pm}
\Big{[}
\Pi_{0\mp bi}\partial_{\pm}x^{(0)i}
+\frac{1}{2}\partial_{\pm} y^{(0)}_{b}
\Big{]}.
\end{eqnarray}
Here
${\tilde{\Theta}}^{ab}_{0\pm}$ 
and $\Pi_{0\mp bi}$
stand for
the zeroth order values of ${\tilde{\Theta}}^{ab}_{\pm}$ and $\Pi_{\mp bi}$,
and they are defined in (\ref{eq:nulte}).

Substituting (\ref{eq:vnulares}) into (\ref{eq:vdef})
we obtain 
\begin{eqnarray}\label{eq:vnula}
&&\Delta V^{(0)a}(x^{i},y_{a})=
\nonumber\\
&&\,
-\kappa\Big[
\tilde\Theta^{ab}_{0+}\Pi_{0- bi}+
\tilde\Theta^{ab}_{0-}\Pi_{0+ bi}
\Big]\Delta x^{(0)i}
\nonumber\\
&&\,
-\kappa\Big[
\tilde\Theta^{ab}_{0+}\Pi_{0- bi}-\tilde\Theta^{ab}_{0-}\Pi_{0+ bi}
\Big]\Delta\tilde{x}^{(0)i}
\nonumber\\
&&\,-\frac{\kappa}{2}\Big[
\tilde\Theta^{ab}_{0+}+\tilde\Theta^{ab}_{0-}
\Big]\Delta y_{b}^{(0)}
-\frac{\kappa}{2}\Big[
\tilde\Theta^{ab}_{0+}-\tilde\Theta^{ab}_{0-}
\Big]\Delta\tilde{y}_{b}^{(0)}.
\nonumber\\
\end{eqnarray}
Here
\begin{eqnarray}\label{eq:tilds}
&&\Delta{\tilde{y}}_{a}^{(0)}=\int_{P}(d\tau y^{(0)\prime}_{a}+d\sigma \dot{y}^{(0)}_{a}),
\nonumber\\
&&\Delta{\tilde{x}}^{(0)i}=\int_{P}(d\tau x^{(0)\prime i}+d\sigma \dot{x}^{(0)i}),
\end{eqnarray}
are the variables T-dual to the coordinates $y_{a}$ and $x^{i}$ in the zeroth order in $B_{\mu\nu\rho}$, for $b_{\mu\nu}=0$, which we call the double variables.

Thus, we obtain the explicit form of the T-dual action and conclude that it is given in terms of the original coordinates $x^{i}$ and the dual coordinates $y_a$ originating from the Lagrange multipliers. However, the background fields' argument depends not only on these variables but on their doubles as well.
Because of this the theory is non-local as the double variables $\tilde{x}^{i}$ and $\tilde{y}_{a}$ are
defined as line integrals.

The action  (\ref{eq:tdualdel}) can be obtained from the initial action (\ref{eq:action1}) under the following substitutions of the coordinate derivatives and the background fields:
\begin{equation}
 \partial_\pm x^i \rightarrow \partial_\pm x^i\, ,\quad \partial_\pm x^a \rightarrow \partial_\pm y_a\, ,
\end{equation}
\begin{eqnarray}
&&\Pi_{+ij}\rightarrow{^\bullet}\Pi_{+ij},\quad
\Pi_{+ia}\rightarrow{^\bullet}\Pi_{+i}^{\ \ \ a},
\nonumber\\
&&\Pi_{+ai}\rightarrow{^\bullet}\Pi^{a}_{+i},\quad
\Pi_{+ab}\rightarrow{^\bullet}\Pi_{+}^{ab},
\end{eqnarray}
where the dual background fields are
\begin{eqnarray}\label{eq:parcijalna}
&&{^\bullet}\Pi_{+ij}=\overline\Pi_{+ij},\quad
{^\bullet}\Pi_{+i}^{\ \ \ a}=-\kappa\Pi_{+ib}\tilde\Theta^{ba}_{-},
\nonumber\\
&&{^\bullet}\Pi^{a}_{+i}=\kappa\tilde\Theta^{ab}_{-}\Pi_{+bi},\quad
{^\bullet}\Pi_{+}^{ab}=\frac{\kappa}{2}\tilde\Theta^{ab}_{-},
\end{eqnarray}
with $\overline\Pi_{+ij}$, $\Pi_{+\mu\nu}$ and $\tilde\Theta^{ab}_{-}$ defined in (\ref{eq:barpi}), (\ref{eq:pi}) and (\ref{eq:barteta1}). The argument of all T-dual background fields is $[x^i,V^a(x^i,y_a)]$.
According to (\ref{eq:vdef}) and (\ref{eq:vnula}),
it is non-local  and consequently non-geometric.
Calculating the symmetric and antisymmetric part of the T-dual field compositions (\ref{eq:parcijalna}),
we find that the T-dual metric and Kalb-Ramond field are
equal to
\begin{eqnarray}\label{eq:tdualnapolja}
&&^\bullet G_{ij}=\overline{G}_{ij}=
G_{ij}
-G_{ia}(\tilde{G}^{-1}_{E})^{ab}G_{bj}
\nonumber\\
&&\quad
-2\kappa\Big(
B_{ia}\tilde\theta^{ab}G_{bj}
+G_{ia}\tilde\theta^{ab}B_{bj}
\Big)
-4B_{ia}(\tilde{G}^{-1}_{E})^{ab}B_{bj},
\nonumber\\
&&^\bullet B_{ij}=\overline{B}_{ij}
=B_{ij}
-\frac{\kappa}{2}G_{ia}\tilde\theta^{ab}G_{bj}
-B_{ia}(\tilde{G}^{-1}_{E})^{ab}G_{bj}
\nonumber\\
&&\quad
-G_{ia}(\tilde{G}^{-1}_{E})^{ab}B_{bj}
-2\kappa B_{ia}\tilde\theta^{ab}B_{bj},
\nonumber\\
&&^\bullet G^{ab}=(\tilde{G}^{-1}_{E})^{ab},
\nonumber\\
&&^\bullet B^{ab}=\frac{\kappa}{2}\tilde\theta^{ab},
\nonumber\\
&&^\bullet G^{a}_{\ i}=
\kappa\tilde\theta^{ab}G_{bi}
+2(\tilde{G}^{-1}_{E})^{ab}B_{bi},
\nonumber\\
&&^\bullet B^{a}_{\ i}=
\kappa\tilde\theta^{ab}B_{bi}
+\frac{1}{2}(\tilde{G}^{-1}_{E})^{ab}G_{bi},
\end{eqnarray}
where $\tilde{G}_{Eab}$ and $\tilde\theta^{ab}$ are defined in (\ref{eq:effmet}) and (\ref{eq:tetamalo}).
The T-dual background fields have the same form as in the flat background
\cite{GPR,B,SN} but in the present case fields  $B_{\mu\nu}$, $\tilde{G}^{-1ab}_{E}$ and $\tilde\theta^{ab}$ are coordinate dependent.

Comparing the solutions for the gauge fields (\ref{eq:vsola}) and (\ref{eq:vres}), we obtain the T-dual coordinate transformation law
\begin{eqnarray}\label{eq:zakonprvi}
&&\partial_{\mp}x^{a}\cong
-2\kappa\tilde{\Theta}^{ab}_{\mp}
\big(x^{i},\Delta V^{a}(x^{i},y_{a})\big)\!\cdot
\nonumber\\
&&\quad\cdot
\Big{[}
\Pi_{\pm bi}\big(x^{i},\Delta V^{a}(x^{i},y_{a})\big)\partial_{\mp}x^{i}
+\frac{1}{2}\partial_\mp y_{b}
\nonumber\\
&&\quad
\mp\beta^{\pm}_{b}\big(x^{i},V^{a}(x^{i},y_{a})\big)
\Big{]}.
\end{eqnarray}


\section{Inverse T-dualization
 ${\cal T}_{a}:S[x^{i},y_{a}]\rightarrow S[x^\mu]$}\label{sec:drugi}
\cleq
In this  section we will show that 
T-dualization of the action $S[x^{i},y_{a}]$,
given by (\ref{eq:tdualdel}),
along already treated directions
$y_{a}$
leads to the original action.

So,
let us localize the global symmetry of the coordinates $y_{a}$
\begin{equation}
\delta y_{a}=\lambda_{a},
\end{equation}
of the action (\ref{eq:tdualdel}).
Note that this is the symmetry, despite the coordinate dependence of the metric (\ref{eq:tdualnapolja}), due to the invariance of the background fields' argument \cite{DS}.
Following the T-dualization procedure,
we substitute the ordinary derivatives with the covariant ones
\begin{equation}
D_\pm y_{a}=\partial_\pm y_{a}+u_{\pm a},
\end{equation}
where $u_{\pm a}$ are gauge fields which transform as $\delta u_{\pm a}=-\partial_\pm \lambda_{a}$. We also substitute coordinates $y_{a}$ in the background fields' argument with the invariant coordinates
\begin{eqnarray}
y^{inv}_{a}&=&\int_{P}(d\xi^{+}D_{+}y_{a}
+d\xi^{-}D_{-}y_{a})
\nonumber\\
&=&
y_{a}(\xi)-y_{a}(\xi_{0})+\Delta U_{a},
\end{eqnarray}
where
\begin{equation}\label{eq:velikou}
\Delta U_{a}=\int_{P}(d\xi^{+}u_{+a}
+d\xi^{-}u_{-a}).
\end{equation}
In this way,
adding the Lagrange multiplier term which makes the introduced gauge fields nonphysical,
we obtain the gauge invariant action
\begin{eqnarray}\label{eq:tdualdelinv}
&&S_{inv}[x^{i},y_{a},y^{inv}_{a},z^{a}]=
\nonumber\\
&&
\kappa\int d^{2}\xi\bigg{[}
\partial_{+}x^{i}{\overline{\Pi}}_{+ij}\big(x^{i},\Delta V^{a}(x^{i},y^{inv}_{a})\big)
\partial_{-}x^{j}
\nonumber\\
&&\qquad-\kappa\,\partial_{+}x^{i}\Pi_{+ia}\big(x^{i},\Delta V^{a}(x^{i},y^{inv}_{a})\big)\!\cdot
\nonumber\\
&&\qquad\quad\cdot
\tilde\Theta^{ab}_{-}\big(x^{i},\Delta V^{a}(x^{i},y^{inv}_{a})\big)D_{-}y_{b}
\nonumber\\
&&\qquad+\kappa\,D_{+}y_{a}\tilde\Theta^{ab}_{-}\big(x^{i},\Delta V^{a}(x^{i},y^{inv}_{a})\big)\!\cdot
\nonumber\\
&&\qquad\quad\cdot
\Pi_{+bi}\big(x^{i},\Delta V^{a}(x^{i},y^{inv}_{a})\big)\partial_{-}x^{i}
\nonumber\\
&&\qquad+\frac{\kappa}{2}\,D_{+}y_{a}\tilde\Theta^{ab}_{-}\big(x^{i},\Delta V^{a}(x^{i},y^{inv}_{a})\big)D_{-}y_{b}
\nonumber\\
&&\qquad+
\frac{1}{2}(u_{+a}\partial_{-}z^{a}-u_{-a}\partial_{+}z^{a})
\bigg{]},
\end{eqnarray} 
which
after fixing the gauge by $y_{a}(\xi)=y_{a}(\xi_{0})$ becomes
\begin{eqnarray}\label{eq:tdualdelinvfix}
&&S_{fix}[x^{i},u_{\pm a},z^{a}]=
\nonumber\\
&&
\kappa\int d^{2}\xi\bigg{[}
\partial_{+}x^{i}{\overline{\Pi}}_{+ij}\big(x^{i},\Delta V^{a}(x^{i},\Delta U_{a})\big)
\partial_{-}x^{j}
\nonumber\\
&&\quad-\kappa\,\partial_{+}x^{i}\Pi_{+ia}\big(x^{i},\Delta V^{a}(x^{i},\Delta U_{a})\big)\!\cdot
\nonumber\\
&&\qquad\cdot
\tilde\Theta^{ab}_{-}\big(x^{i},\Delta V^{a}(x^{i},\Delta U_{a})\big)u_{-b}
\nonumber\\
&&\quad+\kappa\,u_{+a}\tilde\Theta^{ab}_{-}\big(x^{i},\Delta V^{a}(x^{i},\Delta U_{a})\big)\!\cdot
\nonumber\\
&&\qquad\cdot
\Pi_{+bi}\big(x^{i},\Delta V^{a}(x^{i},\Delta U_{a})\big)\partial_{-}x^{i}
\nonumber\\
&&\quad+\frac{\kappa}{2}\,u_{+a}\tilde\Theta^{ab}_{-}\big(x^{i},\Delta V^{a}(x^{i},\Delta U_{a})\big)u_{-b}
\nonumber\\
&&\quad+
\frac{1}{2}(u_{+a}\partial_{-}z^{a}-u_{-a}\partial_{+}z^{a})
\bigg{]},
\end{eqnarray} 
where
$\Delta V^{a}$
is defined in (\ref{eq:vnula}) and
 $\Delta U_{a}$ 
 in (\ref{eq:velikou}).


\subsection{Regaining the T-dual action}

The equations of motion obtained varying the gauge fixed action (\ref{eq:tdualdelinvfix}) over  the Lagrange multipliers $z^{a}$
\begin{equation}\label{eq:lm}
\partial_{+}u_{-a}-\partial_{-}u_{+a}=0,
\end{equation}
have the solution
\begin{equation}\label{eq:vsola1}
u_{\pm a}=\partial_{\pm}y_{a}.
\end{equation}
On this solution the variable $\Delta U_{a}$ defined by (\ref{eq:velikou}) is path independent and reduces to
\begin{equation}
\Delta U_{a}(\xi)=y_{a}(\xi)-y_{a}(\xi_{0}),
\end{equation}
and the gauge fixed action (\ref{eq:tdualdelinvfix}) reduces to the action (\ref{eq:tdualdel}).


\subsection{Regaining the initial action}

The equations of motion obtained varying
the gauge fixed action (\ref{eq:tdualdelinvfix})
over the gauge fields $u_{\pm a}$ are
\begin{eqnarray}\label{eq:ujed}
&&\kappa\tilde\Theta^{ab}_{\mp}\big(x^{i},\Delta V^{a}(x^{i},\Delta U_{a})\big)\cdot
\nonumber\\
&&\cdot
\Big[
\frac{1}{2}u_{\mp b}
+\Pi_{\pm bi}\big(x^{i},\Delta V^{a}(x^{i},\Delta U_{a})\big)\partial_{\mp}x^{i}
\Big]
+\frac{1}{2}\partial_\mp z^{a}=
\nonumber\\
&&
=\pm\kappa\tilde\Theta^{ab}_{0\mp}\beta^{\pm}_{b}\big(x^{i},V^{a}(x^{i},U_{a})\big),
\end{eqnarray}
where terms $\tilde\Theta^{ab}_{0\mp}\beta^{\pm}_{b}$ 
are the contribution from the variation over the  background field argument
\begin{equation}
\delta_{U}S_{fix}=-\kappa^{2}\int d^{2}\xi\big(\delta u_{+a}\tilde\Theta^{ab}_{0-}\beta^{+}_{b}+\delta u_{-a}\tilde\Theta^{ab}_{0+}\beta^{-}_{b}\big).
\end{equation}
Here $\beta^{\pm}_{a}$ is of the same form as (\ref{eq:betaprva})
and $\tilde\Theta^{ab}_{0\mp}$ is defined in (\ref{eq:nulte}).

Let us show that
for the equations of motion (\ref{eq:ujed}),
 the gauge fixed action (\ref{eq:tdualdelinvfix}) will reduce to the initial action (\ref{eq:action1}).
Using the fact that $\tilde\Theta^{ab}_{\mp}$ is inverse to $2\kappa\Pi_{\pm ab}$,
these equations of motion can be rewritten as
\begin{eqnarray}\label{eq:ures}
u_{\mp a}&=&
-2\Pi_{\pm ai}\big(x^{i},\Delta V^{a}(x^{i},\Delta U_{a})\big)
\partial_{\mp}x^{i}
\nonumber\\
&-&2\Pi_{\pm ab}\big(x^{i},\Delta V^{a}(x^{i},\Delta U_{a})\big)
\partial_\mp z^{b}
\nonumber\\
&\pm& 2\beta^{\pm}_{a}\big(x^{i},V^{a}(x^{i},U_{a})\big).
\end{eqnarray}
Substituting (\ref{eq:ures}) into (\ref{eq:tdualdelinvfix}),
using the definition (\ref{eq:barpi}) and the first relation in (\ref{eq:relacije}) one obtains
\begin{eqnarray}\label{eq:iniac}
&&S[x^{i},z^{a}]=
\nonumber\\
&&
\kappa \int_{\Sigma}\! d^2\xi\Big{[}
\partial_{+}x^{i}\Pi_{+ij}\partial_{-}x^{j}
+\partial_{+}x^{i}\Pi_{+ia}\partial_{-}z^{a}
\nonumber\\&&\quad
+\,\partial_{+}z^{a}\Pi_{+ai}\partial_{-}x^{i}
+\partial_{+}z^{a}\Pi_{+ab}\partial_{-}z^{b}
\Big{]}.
\end{eqnarray}
The explicit form of the argument of the background fields
is obtained substituting the
zeroth order of Eq. (\ref{eq:ures})
into (\ref{eq:velikou})
\begin{equation}\label{eq:arg}
U^{(0)}_{a}=-2b_{ai}x^{(0)i}
+G_{ai}\tilde{x}^{(0)i}
-2b_{ab}z^{(0)b}
+G_{ab}\tilde{z}^{(0)b}.
\end{equation}
Consequently, the argument of the background fields $\Delta V^{a}$,
defined in (\ref{eq:vnula}), is just
\begin{equation}\label{eq:Vz}
V^{(0)a}(x^{i},U_{a})=z^{a}.
\end{equation}
Therefore,
the action (\ref{eq:iniac}) is equal to the initial action (\ref{eq:action1}) with $x^\mu=(x^{i},z^{a})$.

Comparing the solutions for the gauge fields (\ref{eq:vsola1})
and (\ref{eq:ures}), we obtain the T-dual transformation law
\begin{eqnarray}\label{eq:zakondrugi}
&&\partial_{\mp}y_{a}\cong
-2\Pi_{\pm ai}(x^{i},z^{a})
\partial_{\mp}x^{i}
-2\Pi_{\pm ab}(x^{i},z^{a})
\partial_\mp z^{b}
\nonumber\\
&&\quad
\pm 2\beta^{\pm}_{a}\big(x^{i},z^{a}\big).
\end{eqnarray}
Substituting $\partial_\mp y_a$ to (\ref{eq:zakonprvi}) with the help of (\ref{eq:Vz}) one finds $\partial_\pm x^{a}=\partial_\pm z^{a}$. So, (\ref{eq:zakondrugi}) is the transformation inverse to (\ref{eq:zakonprvi}), which confirms the relation ${\cal T}^a\circ {\cal T}_a=1$.


\section{T-dualization along all undualized coordinates ${\cal T}^{i}:S[x^{i},y_{a}]\rightarrow\, S[y_\mu]$}\label{sec:treci}
\cleq
In this section we will T-dualize the action (\ref{eq:tdualdel}),
applying the T-dualization procedure to the undualized coordinates $x^{i}$.
Substituting the ordinary derivatives $\partial_\pm x^{i}$ with the
covariant derivatives
\begin{equation}
D_\pm x^{i}=\partial_\pm x^{i}
+w^{i}_\pm,
\end{equation}
where the gauge fields  $w^{i}_\pm$ transform as $\delta w^{i}_\pm=-\partial_\pm\lambda^{i}$, substituting the coordinates $x^{i}$ in the background field arguments by
\begin{equation}
\Delta x^{i}_{inv}=
\int_{P}(d\xi^{+}D_{+}x^{i}
+d\xi^{-}D_{-}x^{i}),
\end{equation}
and adding the Lagrange multiplier term,
we obtain the gauge invariant action
\begin{eqnarray}\label{eq:tdualdelinv1}
&&S_{inv}[x^{i},x^{i}_{inv},y]
=
\nonumber\\
&&
\kappa\int d^{2}\xi\bigg{[}
D_{+}x^{i}{\overline{\Pi}}_{+ij}\big(\Delta x^{i}_{inv},\Delta V^{a}(\Delta x^{i}_{inv},y_{a})\big)
D_{-}x^{j}
\nonumber\\
&&\quad-\kappa\,D_{+}x^{i}\Pi_{+ia}\big(\Delta x^{i}_{inv},\Delta V^{a}(\Delta x^{i}_{inv},y_{a})\big)\!\cdot
\nonumber\\
&&\qquad\cdot
\tilde\Theta^{ab}_{-}\big(\Delta x^{i}_{inv},\Delta V^{a}(\Delta x^{i}_{inv},y_{a})\big)\partial_{-}y_{b}
\nonumber\\
&&\quad+\kappa\,\partial_{+}y_{a}\tilde\Theta^{ab}_{-}\big(\Delta x^{i}_{inv},\Delta V^{a}(\Delta x^{i}_{inv},y_{a})\big)\!\cdot
\nonumber\\
&&\qquad\cdot
\Pi_{+bi}\big(\Delta x^{i}_{inv},\Delta V^{a}(\Delta x^{i}_{inv},y_{a})\big)D_{-}x^{i}
\nonumber\\
&&\quad+\frac{\kappa}{2}\,\partial_{+}y_{a}\tilde\Theta^{ab}_{-}\big(\Delta x^{i}_{inv},\Delta V^{a}(\Delta x^{i}_{inv},y_{a})\big)
\partial_{-}y_{b}
\nonumber\\&&\quad
+\frac{1}{2}(w^{i}_{+}\partial_{-}y_{i}-w^{i}_{-}\partial_{+}y_{i})
\bigg{]}.
\end{eqnarray} 
Substituting the gauge fixing condition $x^{i}(\xi)=x^{i}(\xi_{0})$ one obtains
\begin{eqnarray}\label{eq:actionpfix}
&&S_{fix}[x^{i},w^{i}_\pm,y]=
\nonumber\\
&&\,
\kappa\int d^{2}\xi\bigg{[}
w_{+}^{i}{\overline{\Pi}}_{+ij}
\big(\Delta W\big)
w_{-}^{j}
\nonumber\\
&&\quad
-\kappa\,w_{+}^{i}\Pi_{+ia}
\big(\Delta W\big)
\tilde\Theta^{ab}_{-}
\big(\Delta W\big)
\partial_{-}y_{b}
\nonumber\\
&&\quad+\kappa\,\partial_{+}y_{a}\tilde\Theta^{ab}_{-}\big(\Delta W\big)
\Pi_{+bi}
\big(\Delta W\big)
w_{-}^{i}
\nonumber\\
&&\quad
+\frac{\kappa}{2}\,\partial_{+}y_{a}\tilde\Theta^{ab}_{-}
\big(\Delta W\big)\partial_{-}y_{b}
\nonumber\\
&&\quad+\frac{1}{2}(w^{i}_{+}\partial_{-}y_{i}-w^{i}_{-}\partial_{+}y_{i})
\bigg{]},
\end{eqnarray} 
where $\Delta W^\mu=\big[\Delta W^{i},\Delta V^{a}(\Delta W^{i},y_{a})\big]$
with $\Delta W^{i}$ defined by 
\begin{equation}\label{eq:deltavi}
\Delta W^{i}
\equiv
\int_{P}(d\xi^{+} w^{i}_{+}
+d\xi^{-} w^{i}_{-}),
\end{equation}
and $\Delta V^{a}=\Delta V^{a}(\Delta W^{i},y_{a})$ is defined in (\ref{eq:vnula}), where argument $x^{i}$ is replaced by $\Delta W^{i}$.


\subsection{Regaining the T-dual action}

The equations of motion for the Lagrange multipliers $y_{i}$ are
\begin{equation}\label{eq:emyi}
\partial_{+}w^{i}_{-}
-\partial_{-}w^{i}_{+}=0,
\end{equation}
and they have the solution
\begin{equation}\label{eq:vsola2}
w^{i}_{\pm}=\partial_{\pm}x^{i}.
\end{equation}
For this solution the background field argument $\Delta W^{i}$ defined in (\ref{eq:deltavi})
reduces to
\begin{equation}\label{eq:vvelx2}
\Delta W^{i}(\xi)=x^{i}(\xi)-x^{i}(\xi_{0}),
\end{equation}
so that the argument $\Delta V^{a}$ becomes
\begin{equation}
\Delta V^{a}(\Delta W^{i},y^{a})=\Delta V^{a}(x^{i},y^{a}),
\end{equation}
and therefore the gauge fixed action (\ref{eq:actionpfix}) reduces to the action (\ref{eq:tdualdel}).


\subsection{From the gauge fixed action to the completely T-dual action}

The equations of motion obtained varying the gauge fixed action (\ref{eq:actionpfix}) over $w_\pm^{i}$ are
\begin{eqnarray}\label{eq:jkv}
&&\overline\Pi_{\pm ij}(\Delta W)w^{j}_{\mp}
-\kappa\Pi_{\pm ia}(\Delta W)\tilde\Theta^{ab}_{\mp}(\Delta W)\partial_\mp y_{b}
+\frac{1}{2}\partial_\mp y_{i}
\nonumber\\
&&=\pm2\kappa\overline\Pi_{\pm ij}\Theta^{j\mu}_\mp\beta^\pm_\mu(W),
\end{eqnarray}
where
\begin{equation}\label{eq:betatri}
\beta^\pm_\mu(V)=
\mp \frac{1}{2}h_{\mu\nu}(V)\partial_\mp V^\nu.
\end{equation} 
Terms $\overline\Pi_{\pm ij}\Theta^{j\mu}_\mp\beta^\pm_\mu(W)$ are
the contribution from the background fields argument, defined by 
\begin{eqnarray}
&&\delta_{U}S_{fix}=
\nonumber\\
&&
-2\kappa^{2}\!\int\! d^{2}\xi\Big(\delta w_{+}^{i}\overline\Pi_{+ij}\Theta^{j\mu}_{-}\beta^{+}_{\mu}
+\delta w_{-}^{i}\overline\Pi_{- ij}\Theta^{j\mu}_{+}\beta^{-}_{\mu}\Big),
\nonumber\\
\end{eqnarray}
calculated using (\ref{eq:tetarazvoj}), (\ref{eq:pcrta}), and (\ref{eq:vnula}).

Using the fact that the background field composition $\overline\Pi_{\pm ij}$ is invese to 
$2\kappa\Theta^{ij}_\mp$ defined by (\ref{eq:relacije}), we 
can rewrite the equation of motion (\ref{eq:jkv}) expressing
the gauge fields as
\begin{eqnarray}\label{eq:vresi}
&&w^{i}_{\mp}
=2\kappa\Theta^{ij}_\mp(\Delta W)
\Big{[}
\kappa\Pi_{\pm ja}(\Delta W)\tilde\Theta^{ab}_{\mp}(\Delta W)\partial_\mp y_{b}
\nonumber\\
&&\quad
-\frac{1}{2}\partial_\mp y_{j}
\Big{]}
\pm2\kappa\Theta^{i\mu}_{0\pm}\beta^\pm_\mu(W).
\end{eqnarray}
Using the second relation in (\ref{eq:tetaia}),
we obtain
\begin{eqnarray}\label{eq:vresi1}
w^{i}_{\mp}=
-\kappa\Theta^{i\mu}_\mp(\Delta W)
\Big{[}
\partial_\mp y_\mu
\mp2\beta^\pm_\mu(W)\Big{]}.
\end{eqnarray}

Substituting (\ref{eq:vresi1}) into the gauge fixed action (\ref{eq:actionpfix}), we obtain
\begin{eqnarray}\label{eq:spp}
&&S[y]=
\nonumber\\
&&\kappa\int d^{2}\xi
\bigg{[}
\partial_{+}y_{i}\Big{(}
\kappa\Theta^{ij}_{-}
-\kappa^{2}\Theta^{ik}_{-}\overline\Pi_{+kl}\Theta^{lj}_{-}
\Big{)}\partial_{-}y_{j}
\nonumber\\
&&\quad+\Big{(}
-\kappa^{2}\Theta^{ij}_{-}\overline\Pi_{+jk}\Theta^{ka}_{-}
+\frac{\kappa}{2}\Theta^{ia}_{-}
-\kappa^{2}\Theta^{ij}_{-}\Pi_{+jb}\tilde\Theta^{ba}_{-}
\Big{)}\!\cdot
\nonumber\\
&&\qquad\cdot
\partial_{+}y_{i}\,\partial_{-}y_{a}
\nonumber\\
&&\quad
+\Big{(}
-\kappa^{2}\Theta^{aj}_{-}\overline\Pi_{+jk}\Theta^{ki}_{-}
+\frac{\kappa}{2}\Theta^{ai}_{-}
-\kappa^{2}\tilde\Theta^{ab}_{-}\Pi_{+bj}\Theta^{ji}_{-}
\Big{)}\!\cdot
\nonumber\\
&&\qquad\cdot
\partial_{+}y_{a}\,\partial_{-}y_{i}
\nonumber\\
&&\quad+\partial_{+}y_{a}\Big{(}
\frac{\kappa}{2}\tilde\Theta^{ab}_{-}
-\kappa^{2}\Theta^{ai}_{-}\overline\Pi_{+ij}\Theta^{jb}_{-}
\nonumber\\
&&\qquad
-\kappa^{2}\Theta^{ai}_{-}\Pi_{+ic}\tilde\Theta^{cb}_{-}
-\kappa^{2}\tilde\Theta^{ac}_{-}\Pi_{+ci}\Theta^{ib}_{-}
\Big{)}
\partial_{-}y_{b}
\bigg{]}.
\end{eqnarray}
Using (\ref{eq:relacije}), (\ref{eq:rel1}), and (\ref{eq:tbpttp})
one can rewrite this action as
\begin{equation}\label{eq:tdualna}
S[y]=\frac{\kappa^{2}}{2}\int d^{2}\xi\,
\partial_{+}y_\mu\Theta^{\mu\nu}_{-}\big(\Delta W\big)\partial_{-}y_\nu.
\end{equation}

In order to find the background fields argument $\Delta W^{i}$,
we consider the zeroth order of Eqs. (\ref{eq:vresi1}),
and conclude that
\begin{equation}\label{eq:argtri}
\Delta W^{i}=-\kappa\theta_{0}^{i\mu}\Delta y_\mu+(g^{-1})^{i\mu}\Delta{\tilde{y}}_\mu.
\end{equation}
Using (\ref{eq:pomoc}) and (\ref{eq:tetaia}),
we find that
$\Delta V^{a}(\Delta W^{i},y^{a})$ defined in
(\ref{eq:vnula}) equals
\begin{equation}\label{eq:argtrid}
\Delta V^{a}(\Delta W^{i},y_{a})=
-\kappa\theta_{0}^{a\mu}\Delta y_\mu+(g^{-1})^{a\mu}\Delta{\tilde{y}}_\mu.
\end{equation}
Therefore,
we conclude that the background fields' argument is equal to (\ref{eq:vargdef}),
so that the action (\ref{eq:tdualna}) is the completely T-dual action (\ref{eq:dualna}),
which is in agreement with Ref. \cite{DS}.
Comparing the solutions for the gauge fields (\ref{eq:vsola2}) and (\ref{eq:vresi1}), we obtain the T-dual transformation law
\begin{equation}\label{eq:zakontreci}
\partial_{\mp}x^{i}
\cong
-\kappa\Theta^{i\mu}_\mp\big(\Delta V(y)\big)\Big[
\partial_\mp y_\mu
\mp2\beta^\pm_{\mu}\big(V(y)\big)
\Big].
\end{equation}

One can verify that two successive T-duality transformations (\ref{eq:zakonprvi}) and (\ref{eq:zakontreci}) 
correspond to the total T-duality transformation (\ref{eq:zak}). Indeed, the relation (\ref{eq:zakontreci}) is just the $i$th component of this transformation. Substituting $\partial_\pm x^i$ from (\ref{eq:zakontreci}) into (\ref{eq:zakonprvi}),
using (\ref{eq:bttt}) and (\ref{eq:tbpttp}), we obtain
$$\partial_\pm x^a=-\kappa\Theta^{a\mu}_\pm\big(\Delta V\big)
\Big{[}
\partial_\pm y_\mu
\pm2\beta^\mp_\mu\big(V\big)
\Big{]},
$$
which is just the $a$th component of the complete T-duality transformation. So, we confirm that ${\cal T}^a\circ{\cal T}^i={\cal T}$.


\section{Inverse T-dualization along arbitrary subset of the dual coordinates ${\cal T}_{i}:S[y_\mu]\rightarrow S[x^{i},y_{a}]$ }\label{sec:cetvrti}
\cleq

Finally,
in this section we will show that the T-du\-a\-li\-za\-ti\-on of the completely T-du\-a\-l action
(\ref{eq:dualna}), along arbitrary subset of the dual  coordinates $y_{i}$ leads to T-dual action (\ref{eq:tdualdel}).
So, let us start with the T-dual action
\begin{equation}\label{eq:dualnapon}
S[y]=\frac{\kappa^{2}}{2}
\int d^{2}\xi\
\partial_{+}y_\mu
\Theta_{-}^{\mu\nu}\big(\Delta V(y)\big)
\partial_{-}y_\nu,
\end{equation}
which is globally invariant to the constant shift of coordinates $y_\mu$
\begin{equation}
\delta y_\mu=\lambda_\mu.
\end{equation}
We localize this symmetry for the coordinates $y_{i}$ and obtain
the locally invariant action
\begin{eqnarray}
&&S_{inv}[y,y_{i}^{inv},z^{i}]=
\nonumber\\
&&
\frac{\kappa^{2}}{2}
\int d^{2}\xi\Big{[}
D_{+}y_{i}
\Theta_{-}^{ij}\big(\Delta V(y_{i}^{inv},y_{a})\big)
D_{-}y_{j}
\nonumber\\
&&\quad
+D_{+}y_{i}
\Theta_{-}^{ia}\big(\Delta V(y_{i}^{inv},y_{a})\big)
\partial_{-}y_{a}
\nonumber\\
&&\quad
+\partial_{+}y_{a}
\Theta_{-}^{ai}\big(\Delta V(y_{i}^{inv},y_{a})\big)
D_{-}y_{i}
\nonumber\\
&&\quad
+\partial_{+}y_{a}
\Theta_{-}^{ab}\big(\Delta V(y_{i}^{inv},y_{a})\big)
\partial_{-}y_{b}
\nonumber\\
&&\quad
+\frac{1}{\kappa}(u_{+i}\partial_{-}z^{i}-u_{-i}\partial_{+}z^{i})
\Big{]},
\end{eqnarray}
where
$D_\pm y_{i}=\partial_\pm y_{i}+u_{\pm i}$ are the covariant derivatives.
The gauge fields $u_{\pm i}$ transform as $\delta u_{\pm i}=-\partial_\pm\lambda_{i}$  and
the invariant coordinates are defined by $y_{i}^{inv}=\int_{P}(d\xi^{+}D_{+}y_{i}
+d\xi^{-}D_{-}y_{i})$.
After fixing the gauge by
$y_{i}(\xi)=y_{i}(\xi_{0})$, the action becomes
\begin{eqnarray}\label{eq:sfixdual}
&&S_{fix}[y_{a},u_{\pm i},z^{i}]=
\nonumber\\
&&\frac{\kappa^{2}}{2}
\int d^{2}\xi\Big{[}
u_{+i}
\Theta_{-}^{ij}\big(\Delta V(\Delta U_{i},y_{a})\big)
u_{-j}
\nonumber\\
&&\quad
+u_{+i}
\Theta_{-}^{ia}\big(\Delta V(\Delta U_{i},y_{a})\big)
\partial_{-}y_{a}
\nonumber\\
&&\quad+
\partial_{+}y_{a}
\Theta_{-}^{ai}\big(\Delta V(\Delta U_{i},y_{a})\big)
u_{-i}
\nonumber\\
&&\quad+
\partial_{+}y_{a}
\Theta_{-}^{ab}\big(\Delta V(\Delta U_{i},y_{a})\big)
\partial_{-}y_{b}
\nonumber\\
&&\quad+\frac{1}{\kappa}(u_{+i}\partial_{-}z^{i}-u_{-i}\partial_{+}z^{i})
\Big{]},
\end{eqnarray}
where $\Delta U_{i}=\int_{P}(d\xi^{+}u_{+i}
+d\xi^{-}u_{-i})$.


\subsection{Regaining the T-dual action}

The equations of motion obtained varying the gauge fixed action (\ref{eq:sfixdual})
over the Lagrange multipliers
\begin{equation}
\partial_{+}u_{-i}-
\partial_{-}u_{+i}=0,
\end{equation}
have the solution
\begin{equation}\label{eq:vsola3}
u_{\pm i}=\partial_{\pm}y_{i}.
\end{equation}
On this solution the variable $\Delta U_{i}$
reduces to
\begin{equation}\label{eq:vvelx3}
\Delta U_{i}(\xi)=y_{i}(\xi)-y_{i}(\xi_{0}),
\end{equation}
and therefore
\begin{equation}
\Delta V^\mu(\Delta U_{i},y_{a})=\Delta V^\mu(y).
\end{equation}
So, the action (\ref{eq:sfixdual}) becomes the action (\ref{eq:dualnapon}).


\subsection{Obtaining the T-dual action}

The equations of motion obtained varying the action (\ref{eq:sfixdual}) over $u_{\pm i}$ are
\begin{eqnarray}\label{eq:ujk}
&&\kappa\Theta^{ij}_\mp\big(\Delta V(\Delta U_{i},y_{a})\big) u_{\mp j}
+\kappa\Theta^{ia}_\mp\big(\Delta V(\Delta U_{i},y_{a})\big)\partial_\mp y_{a}
\nonumber\\
&&
+\partial_\mp z^{i}
=\pm2\kappa\Theta^{i\mu}_{0\mp}\beta^{\pm}_{\mu}\big(V(U_{i},y_{a})\big),
\end{eqnarray}
where $\beta^{\pm}_\mu$ are 
given by (\ref{eq:betatri}).
The terms with beta function
come from the variation over the argument $U_{i}$
\begin{equation}
\delta_{U}S_{fix}=-{\kappa^{2}}\int d^{2}\xi\big(\delta u_{+i}\Theta^{i\mu}_{0-}\beta^{+}_\mu+\delta u_{-i}\Theta^{i\mu}_{0+}\beta^{-}_\mu\big),
\end{equation}
and are calculated
using (\ref{eq:tetarazvoj}) and (\ref{eq:vargdef}).
Using the fact that $2\kappa\overline\Pi_{\mp ij}$ is the inverse of $\Theta^{ij}_\pm$,
the equation (\ref{eq:ujk}) can be rewritten as
\begin{eqnarray}\label{eq:uresi}
&&u_{\mp i}=
-2\overline\Pi_{\pm ij}\big(\Delta V(\Delta U_{i},y_{a})\big)
\Big{[}
\kappa\Theta^{ja}_\mp\big(\Delta V(\Delta U_{i},y_{a})\big)\!\cdot
\nonumber\\
&&\cdot
\partial_\mp y_{a}
+\partial_\mp z^{j}
\mp2\kappa\Theta^{j\mu}_{0\mp}\beta^{\pm}_\mu\big(V(U_{i},y_{a})\big)
\Big{]}.
\end{eqnarray}
Substituting (\ref{eq:uresi}) into the gauge fixed action (\ref{eq:sfixdual}),
using (\ref{eq:bttt}) we obtain
\begin{eqnarray}
&&S[z^{i},y_{a}]=
\nonumber\\
&&\frac{\kappa^{2}}{2}\int d^{2}\xi
\Big{[}
\frac{2}{\kappa}\partial_{+}z^{i}\overline\Pi_{+ij}\partial_{-}z^{j}
+2\partial_{+}z^{i}\overline\Pi_{+ij}\Theta^{jb}_{-}\partial_{-}y_{b}
\nonumber\\
&&\quad-2\partial_{+}y_{a}\Theta^{ai}_{-}\overline\Pi_{+ij}\partial_{-}z^{j}
+\partial_{+}y_{a}\tilde\Theta^{ab}_{-}\partial_{-}y_{b}
\Big{]},
\end{eqnarray}
which with the help of (\ref{eq:tbpttp}) becomes
\begin{eqnarray}\label{eq:pdd}
&&S[z^{i},y_{a}]=\frac{\kappa^{2}}{2}\int d^{2}\xi
\Big{[}
\frac{2}{\kappa}\partial_{+}z^{i}\overline\Pi_{+ij}\partial_{-}z^{j}
\nonumber\\
&&
\quad-2\partial_{+}z^{i}\Pi_{+ia}\tilde\Theta^{ab}_{-}\partial_{-}y_{b}
+\,2\partial_{+}y_{a}\tilde\Theta^{ab}_{-}\Pi_{+bj}\partial_{-}z^{j}
\nonumber\\
&&\quad
+\partial_{+}y_{a}\tilde\Theta^{ab}_{-}\partial_{-}y_{b}
\Big{]}.
\end{eqnarray}

In order to find
the argument of the background fields $\Delta V(\Delta U_{i},y_{a})$,
one considers the
zeroth order of the equations (\ref{eq:uresi}) and obtains
\begin{eqnarray}\label{eq:unula}
\Delta U^{(0)}_{i}&=&
-\Big[
\overline\Pi_{0+ij}+\overline\Pi_{0-ij}
\Big]\Delta z^{(0)j}
\nonumber\\
&+&\Big[\overline\Pi_{0+ij}-\overline\Pi_{0-ij}
\Big]\Delta \tilde{z}^{(0)j}
\nonumber\\
&-&\kappa\Big[
\overline\Pi_{0+ij}\Theta_{0-}^{ja}+
\overline\Pi_{0-ij}\Theta_{0+}^{ja}
\Big]\Delta y^{(0)}_{a}
\nonumber\\
&+&\kappa\Big[
\overline\Pi_{0+ij}\Theta_{0-}^{ja}-
\overline\Pi_{0-ij}\Theta_{0+}^{ja}
\Big]\Delta\tilde{y}^{(0)}_{a},
\end{eqnarray}
where the double variables are defined in analogy with (\ref{eq:tilds}).
Substituting (\ref{eq:unula})
into (\ref{eq:vargdef}),
we obtain
\begin{equation}
\Delta V^{i}(\Delta U_{i},y_{a})=
\Delta z^{i},
\end{equation}
and
\begin{eqnarray}
\Delta V^{a}(\Delta U_{i},y_{a})
&=&-\kappa\Big[
\tilde\Theta^{ab}_{0+}\Pi_{0- bi}+
\tilde\Theta^{ab}_{0-}\Pi_{0+ bi}
\Big]\Delta z^{(0)i}
\nonumber\\
&-&\kappa\Big[
\tilde\Theta^{ab}_{0+}\Pi_{0- bi}-\tilde\Theta^{ab}_{0-}\Pi_{0+ bi}
\Big]\Delta\tilde{z}^{(0)i}
\nonumber\\
&-&\frac{\kappa}{2}\Big[
\tilde\Theta^{ab}_{0+}+\tilde\Theta^{ab}_{0-}
\Big]\Delta y_{b}^{(0)}
\nonumber\\
&-&\frac{\kappa}{2}\Big[
\tilde\Theta^{ab}_{0+}-\tilde\Theta^{ab}_{0-}
\Big]\Delta\tilde{y}_{b}^{(0)},
\end{eqnarray}
which is exactly (\ref{eq:vnula}) with $z^{i}=x^{i}$. So, we can conclude that the action (\ref{eq:pdd}) is equal to the T-dual
action (\ref{eq:tdualdel}).

Comparing the solutions for the gauge fields (\ref{eq:vsola3}) and (\ref{eq:uresi}),
we obtain the T-dual transformation law
\begin{eqnarray}\label{eq:zakoncetiri}
&&\partial_{\mp}y_{i}\cong
-2\overline\Pi_{\pm ij}
\big(\Delta z^{i},\Delta V^{a}
(\Delta U_{i}(z^{i},y_{a}),y_{a})\big)\!\cdot
\nonumber\\
&&\quad\cdot
\Big{[}
\kappa\Theta^{ja}_\mp
\big(\Delta z^{i},\Delta V^{a}
(\Delta U_{i}(z^{i},y_{a}),y_{a})\big)
\partial_\mp y_{a}
+\partial_\mp z^{j}
\nonumber\\
&&\quad
\mp2\kappa\Theta^{j\mu}_{0\mp}\beta^{\pm}_\mu
\big( z^{i},V^{a}(U_{i}(z^{i},y_{a}),y_{a})\big)
\Big{]}.
\end{eqnarray}
These transformations are inverse to (\ref{eq:zakontreci}), so that ${\cal T}^i\circ {\cal T}_i=1$.
Successively applying (\ref{eq:zakoncetiri}) and (\ref{eq:zakondrugi}),
using (\ref{eq:tbpttp}) and (\ref{eq:bttt}),
 we obtain the $i$th component of the inverse law of the total T-dualization (\ref{eq:zakinv}).
Its $a$th component is (\ref{eq:zakondrugi}), so we confirm that ${\cal T}_{a}\circ {\cal T}_{i}=\tilde{\cal T}$. 


\section{Group of the T-dual transformation laws}
\cleq
In this section we will recapitulate the coordinate transformation laws between the theories considered.
In Sect. \ref{sec:prvi}, we performed T-dualization procedure along coordinates $x^{a}$
\begin{equation}
{\cal T}^{a}:S[x^\mu]\rightarrow S[x^{i},y_{a}],
\end{equation} 
and obtained the following coordinate transformation law (\ref{eq:zakonprvi})
\begin{eqnarray}\label{eq:p}
&&\partial_{\mp}x^{a}\cong
-2\kappa\tilde{\Theta}^{ab}_{\mp}
\big(x^{i},\Delta V^{a}(x^{i},y_{a})\big)\cdot
\nonumber\\
&&\cdot
\Big{[}
\Pi_{\pm bi}\big(x^{i},\Delta V^{a}(x^{i},y_{a})\big)\partial_{\mp}x^{i}
+\frac{1}{2}\partial_\mp y_{b}
\nonumber\\
&&\mp\beta^{\pm}_{b}\big(x^{i},V^{a}(x^{i},y_{a})\big)
\Big{]}
\end{eqnarray} 
where $V^{a}$ and $\beta^{\pm}_{a}$ are given by (\ref{eq:vnula}) and (\ref{eq:betaprva}).
In the zeroth oder this law implies
\begin{equation}\label{eq:kor}
x^{(0)a}\cong V^{(0)a}(x^{i},y_{a}).
\end{equation}

In Sect. \ref{sec:drugi}, starting from the action $S[x^{i},y_{a}]$ we performed
T-dualization procedure along coordinates $y_{a}$
\begin{equation}
{\cal T}_{a}:S[x^{i},y_{a}]\rightarrow S[x^\mu],
\end{equation}
and obtained the transformation law (\ref{eq:zakondrugi})
\begin{equation}\label{eq:d}
\partial_{\mp}y_{a}\cong
-2\Pi_{\pm a\mu}(x)
\partial_{\mp}x^\mu
\pm 2\beta^{\pm}_{a}(x),
\end{equation}
which is the law inverse to (\ref{eq:p})
and in the zeroth order it implies
\begin{equation}
y^{(0)}_{a}\cong U^{(0)}_{a}(x).
\end{equation}
Multiplying the transformation law (\ref{eq:p}) from the left side by $\Pi_{\pm ca}(x)\cong\Pi_{\pm ca}\big(x^{i},\Delta V^{a}(x^{i},y_{a})\big)$, using (\ref{eq:kor}),
we obtain the transformation law (\ref{eq:d}). So, we confirm that ${\cal T}^{a}\circ{\cal T}_{a}=1$.

In Sect. \ref{sec:treci},
starting once again from the action $S[x^{i},y_{a}]$,
we performed T-dualization procedure along the undualized coordinates $x^{i}$
\begin{equation}
{\cal T}^{i}:S[x^{i},y_{a}]\rightarrow\, S[y_\mu],
\end{equation}
and obtained the coordinate transformation law (\ref{eq:zakontreci})
\begin{equation}\label{eq:t}
\partial_{\mp}x^{i}
\cong
-\kappa\Theta^{i\mu}_\mp\big(\Delta V(y)\big)\Big[
\partial_\mp y_\mu
\mp2\beta^\pm_{\mu}\big(V(y)\big)
\Big],
\end{equation}
where $V^\mu$ and $\beta^\pm_{\mu}$ are given by (\ref{eq:vargdef})  and  (\ref{eq:betatri}).
In the zeroth order it gives
\begin{equation}\label{eq:korr}
x^{(0)i}\cong V^{(0)i}(y).
\end{equation}
The two successive T-duality transformations (\ref{eq:p}) and (\ref{eq:t}) give the complete transformation (\ref{eq:zak}),
so that ${\cal T}^{a}\circ{\cal T}^{i}={\cal T}$.

In Sect. \ref{sec:cetvrti},
starting from the completely T-dual action $S[y]$,
we performed T-dualization procedure along coordinates $y_{i}$
\begin{equation}
{\cal T}_{i}:S[y_\mu]\rightarrow S[x^{i},y_{a}],
\end{equation}
and obtained (\ref{eq:zakoncetiri})
\begin{eqnarray}\label{eq:c}
&&\partial_{\mp}y_{i}\cong
-2\overline\Pi_{\pm ij}
\big(\Delta x^{i},\Delta V^{a}
(\Delta U_{i}(x^{i},y_{a}),y_{a})\big)\cdot
\nonumber\\
&&\quad\cdot
\Big{[}
\kappa\Theta^{ja}_\mp
\big(\Delta x^{i},\Delta V^{a}
(\Delta U_{i}(x^{i},y_{a}),y_{a})\big)
\partial_\mp y_{a}
+\partial_\mp x^{j}
\nonumber\\
&&\quad
\mp2\kappa\Theta^{j\mu}_{0\mp}\beta^{\pm}_\mu
\big( x^{i},V^{a}(U_{i}(x^{i},y_{a}),y_{a})\big)
\Big{]},
\end{eqnarray}
with $V^{a}$, $U_{i}$ and $\beta^{\pm}_\mu$ given by (\ref{eq:argtrid}), (\ref{eq:unula}) and (\ref{eq:betatri}).
In the zeroth order this law implies
\begin{equation}
y^{(0)}_{i}\cong U^{(0)}_{i}(x^{i},y_{a}).
\end{equation}
Multiplying  (\ref{eq:c}) from the left by $$\Theta^{ki}_\mp\big(\Delta x^{i},\Delta V^{a}(y)\big)\cong\Theta^{ki}_\mp\big(\Delta x^{i},\Delta V^{a}
(\Delta U_{i}(x^{i},y_{a}),y_{a})\big),$$ using (\ref{eq:korr}), we obtain the transformation law (\ref{eq:t}),
so that ${\cal T}^{i}\circ{\cal T}_{i}=1.$
Successively applying (\ref{eq:c}) and (\ref{eq:d}),
using (\ref{eq:tbpttp}) and (\ref{eq:bttt}),
 we obtain the $i$th component of the inverse law of the complete T-dualization (\ref{eq:zakinv}).
Its $a$th component is (\ref{eq:d}), so we confirm that ${\cal T}_{a}\circ {\cal T}_{i}=\tilde{\cal T}$.

We can conclude that the
elements $1,{\cal T}^{a}$ and ${\cal T}_{a}$, with $d=1,\dots,D$, 
form an Abelian group.
The element ${\cal T}^{a}$ is the inverse of the element ${\cal T}_{a}$.
\section{Dilaton field in the weakly curved background}
\cleq
The  T-duality transformation of the dilaton field in the weakly curved background was considered in Ref.\cite{BS}. For completeness and  further use,  we give here a brief recapitulation of some basic steps of the treatment.

It is well known that  a dilaton transformation has a quantum origin.
So, let us start with the path integral for the gauge fixed action
\begin{equation}\label{eq:fi}
{\cal Z} = \int d v_+^\mu   d v_-^\mu dy_\mu    e^{i S_{fix} (v_{\pm}, \partial_\pm y)}   \, ,
\end{equation}
where
\begin{equation}
S_{fix} (v_{\pm}, \partial_\pm y) = S_0 + S_1 \,  ,              
\end{equation}
with $S_1$ being the infinitesimal part of the action
\begin{eqnarray}
&& S_0 = \kappa \int  d^2 \xi [v_+^\mu \Pi_{0+\mu\nu}  v_-^\nu + \frac{1}{2} (v_+^\mu \partial_- y_\mu - v_-^\mu \partial_+ y_\mu ) ] \,  ,
\nonumber\\
&&
S_1 = \kappa \int  d^2 \xi \, v_+^\mu h_{\mu\nu}(V) v_-^\nu   \,  .
\end{eqnarray}

For a constant background ($S_1=0$) the path integral is  Gaussian and it  equals $(\det \Pi_{0 + \mu \nu})^{-1}$. In our case the background is coordinate dependent and thus the integral is not Gaussian. The fact that we work with an infinitesimal parameter enables us to show that the final result is formally the same as in the flat case \cite{BS}
\begin{equation}\label{eq:fi4}
{\cal Z} = \int dy_\mu    \, \frac{1}{det(\Pi_{+\mu\nu}(V))}   \,   e^{i \, {}^\star S (y) }   \, ,
\end{equation}
where ${}^\star S (y)= \frac{\kappa^2}{2} \int d^2 \xi \, \partial_+ y_\mu \, \Theta_{-}^{\mu\nu}(V)  \partial_- y_\nu$ is the complete T-dual action and $\Pi_{+\mu\nu}(V)=B_{\mu\nu}(V)+\frac{1}{2}G_{\mu\nu}$.
Consequently, although for the weakly curved background the functional integration over $v_\pm$ is of the third degree, it produces formally the same result as in the flat space
(where the action is Gaussian),
\begin{equation}\label{eq:diltrans}
{}^\bullet \Phi=\Phi-\ln\det\sqrt{2\Pi_{+ab}}\, .
\end{equation}

Using the expressions for T-dual fields  (\ref{eq:parcijalna})    we can find the relations between the determinants
\begin{eqnarray}\label{eq:dr}
&&\det (2 \Pi_{\pm ab})=  \frac{1}{\det ( 2 \,{^\bullet \Pi}_{\pm}^{ ab})} = \sqrt{\frac{\det G_{a b}}{\det {^\bullet G}^{a b}}}
\nonumber\\
&&\quad
 = \sqrt{\frac{\det G_{\mu \nu}}{\det{^\bullet G}_{\mu \nu}}}  \, ,
\end{eqnarray}
where because of the relation $\Pi_{\pm ab} = B_{a b}\pm \frac{1}{2}G_{a b}$  we put in the factor $2$ for convenience. The symbol ${}^\bullet G_{\mu\nu}$ denotes metric in the whole space-time after partial
T-dualization along $x^a$ directions.
With the help of last relation we can show that the change of space-time measure in the path integral is correct
\begin{eqnarray}\label{eq:ccm}
&&\sqrt{\det G_{\mu \nu}} \, d x^i d x^a  \to     \sqrt{\det G_{\mu \nu}}\,  d x^i    \frac{1}{\det ( 2 \,  \Pi_{+ ab})}\,  d y_a 
\nonumber\\
&&
=
\sqrt{\det {^\bullet G}_{\mu \nu}} \, d x^i d y_a \, ,
\end{eqnarray}
when we performed T-dualization $T^a$ along  $x^a$ directions.


\section{Comparison with the existing facts}
\cleq
\subsection{T-dualization chain for the background with $H$ flux}

In this section we will compare
our results with the T-du\-ali\-za\-tion chain of Ref. \cite{ALLP}.
The coordinates of the $D=3$-dimensional torus will be denoted by $x^{1},x^{2},x^{3}$.
Because of the different notation,
the background fields considered in this paper
and those considered in \cite{ALLP},
which will be denoted ${\cal G}$ and ${\cal B}$, are related by
\begin{equation}
{\cal B}_{\mu\nu}=-2B_{\mu\nu}\, ,\quad {\cal G}_{\mu\nu}=G_{\mu\nu},
\quad\mu,\nu=1,2,3.
\end{equation}

Nontrivial components of the  background considered in
Ref.\cite{ALLP} are
\begin{equation}\label{eq:pp}
{\cal G}_{\mu\nu}=\delta_{\mu\nu}\, ,\quad {\cal B}_{12}=H x^3\, , 
\end{equation}
which in our notation corresponds to the background fields
\begin{equation}\label{eq:nasapolja}
G_{\mu\nu}=\delta_{\mu\nu}\, ,\quad B_{12}=-\frac{1}{2}Hx^3.
\end{equation}

Let us first compare the results in the case $d=1$,
corresponding to the transition
$$ {\it \,T^{1}: torus\, with\, H\scalebox{0.75}[1.0]{\( - \)}flux \rightarrow twisted\, torus}.
$$
To do so,
let us perform T-dualization along direction $x^{1}$,
$T^{1}:S[x]\rightarrow S[y_{1},x^{2},x^{3}]$,
for the string moving in the background (\ref{eq:nasapolja}).
The indices take the values
$a,b\in\{1\}$ and $i,j\in\{2,3\}$.
Because the only nontrivial component of the Kalb-Ramond field is $B_{ai}=-\frac{1}{2}Hx^{3}\delta_{i2}\,$,
the effective fields are just $\tilde{G}^{E}_{\mu\nu}=\delta_{\mu\nu}$
and $\tilde\theta^{ab}=0$.
So, the T-dual background fields (\ref{eq:tdualnapolja}), in the linear order in $H$, are 
\begin{eqnarray}
&^\bullet G_{ij}=\delta_{ij},
&\quad
^\bullet B_{ij}=0,
\nonumber\\
&^\bullet G^{ab}=\delta^{ab},
&\quad
^\bullet B^{ab}=0,
\nonumber\\
&^\bullet G^a{}_i=-Hx^3\delta_{i2},
&\quad
^\bullet B^a{}_i=0.
\end{eqnarray}
Therefore
\begin{equation}
{}^\bullet G_{\mu\nu}=\left(
\begin{array}{ccc}
1 & -Hx^3 & 0\\
-Hx^3 & 1 & 0\\
0 & 0 & 1
\end{array}
\right)=\,^\bullet{\cal G}_{\mu\nu},
\end{equation}
and
\begin{equation}
{}^\bullet B_{\mu\nu}=0={}^\bullet {\cal B}_{\mu\nu},
\end{equation}
so our result is in agreement with that of Ref. \cite{ALLP}.

Now, let us make the comparison in the case
$d=2$ which corresponds to the transition
$$T^{1}\circ T^{2}:{\it torus\, with\, H\scalebox{0.75}[1.0]{\( - \)} flux
 \rightarrow Q\scalebox{0.75}[1.0]{\( - \)}flux\, non\scalebox{0.75}[1.0]{\( - \)}geometry}.$$
Instead to perform $T^{2}$ dualization, from twisted torus to $Q$-flux non-geometry as in \cite{ALLP},
we will start from the initial background with $H$-flux and perform
T-du\-a\-li\-za\-ti\-ons along $x^{1}$ and $x^{2}$,
$T^{1}\circ T^{2}:S[x]\rightarrow S[y_{1},y_{2},x^{3}]$.
The indices take the values
$a,b\in\{1,2\}$ and $i,j\in\{3\}$. Because the only nontrivial contribution to the Kalb-Ramond field $B_{ab}$ is $B_{12}=-\frac{1}{2}Hx^3$, 
the effective background fields are
$\tilde{G}^{E}_{ab}=\delta_{ab},\,\bar{G}^{E}_{ij}=\delta_{ij}$
and the only nonzero component of $\tilde\theta^{ab}$ is $\tilde\theta^{12}=\frac{1}{\kappa}Hx^{3}$.
The T-dual background fields linear in $H$ are therefore
\begin{equation}
{}^\bullet G_{ij}=\delta_{ij}\, ,\quad {}^\bullet G^{ab}=\delta^{ab}\, ,\quad {}^\bullet G^a{}i=0,
\end{equation}
and
\begin{equation}
{}^\bullet B_{ij}=0\, ,\quad {}^\bullet B^{12}=\frac{1}{2}Hx^3\, ,\quad {}^\bullet B^a{}_i=0.
\end{equation}
Consequently
\begin{eqnarray}
{}^\bullet G_{\mu\nu}=\left(
\begin{array}{ccc}
1 & 0 & 0\\
0 & 1 & 0\\
0 & 0 & 1
\end{array}
\right)=\,^\bullet{\cal G}_{\mu\nu},
\end{eqnarray}
\begin{eqnarray}
^\bullet {\cal B}_{\mu\nu}=-2\,^\bullet B_{\mu\nu}=\left(
\begin{array}{ccc}
0 & -Hx^3 & 0\\
Hx^3 & 0 & 0\\
0 & 0 & 0
\end{array}
\right)\, ,
\end{eqnarray}
so the results of this paper and \cite{ALLP} in this case coincide.

\subsection{Non-associativity of R-flux background and breaking of Jacobi identity}

In Ref. \cite{DS,DNS} we obtained T-dual transformation laws connecting T-dual coordinates $y_\mu$ with the initial coordinates $x^\mu$.
Here we will reduce our case to the 3-dimensional torus with H-flux considered in \cite{BDLPR}. Then, the full T-dualization along all coordinates
corresponds to the so-called R-flux. So, we are going to calculate its characteristic features: nonassociativity relation and breaking of Jacobi identity.

We will work in the background of Sect. 9.1 consisting of euclidean flat metric $G_{\mu\nu}$ and Kalb-Ramond field with one nontrivial component $B_{12}=-\frac{1}{2}Hx^3$. T-dual transformation laws for coordinates $y_\mu \;(\mu=1,2,3)$ are of 
the form
\begin{equation}\label{eq:jed1}
y'_1 \cong \frac{1}{\kappa}\pi_1+\frac{1}{2}Hx^3 x'^2\, ,
\end{equation}
\begin{equation}\label{eq:jed2}
y'_2 \cong \frac{1}{\kappa}\pi_2-\frac{1}{2}Hx^3 x'^1\, ,
\end{equation}
\begin{equation}\label{eq:jed3}
y'_3 \cong \frac{1}{\kappa}\pi_3\, ,
\end{equation}
where $\pi_1\, ,\pi_2\, \pi_3$ are canonically conjugated momenta for coordinates $x^1\, ,x^2\, ,x^3$, respectively. Initial space is geometric one, so, the standard Poisson algebra is satisfied 
\begin{eqnarray}\label{eq:spa}
&&\{x^\mu(\sigma),\pi_\nu(\bar\sigma)\}=\delta^\mu{}_\nu\delta(\sigma-\bar\sigma)\, ,
\nonumber\\ &&\{x^\mu,x^\nu\}=\{\pi_\mu,\pi_\nu\}=0\, .
\end{eqnarray}
From (\ref{eq:jed1})-(\ref{eq:jed3}) we obtain
\begin{equation}
\{y'_\mu(\sigma),y'_\nu(\bar\sigma)\}=-\frac{1}{2\kappa}H\varepsilon_{\mu\nu\rho}x'^\rho \delta(\sigma-\bar\sigma)\, ,
\end{equation}
which, after two partial integrations, produces
\begin{equation}
\{y_\mu(\sigma),y_\nu(\bar\sigma)\}=\frac{1}{2\kappa}H\varepsilon_{\mu\nu\rho}\left[x^\rho(\sigma)-x^\rho(\bar\sigma)\right] \theta(\sigma-\bar\sigma)\, ,
\end{equation}
where $\varepsilon_{\mu\nu\rho}$ is 3-dimensional Levi-Civita tensor ($\varepsilon_{123}=1$) and the function $\theta(\sigma)$ is defined as
\begin{equation}\label{eq:fdelt}
\theta(\sigma)\equiv
\left\{\begin{array}{ll}
0 & \textrm{if $\sigma=0$}\\
1/2 & \textrm{if $0<\sigma<2\pi$}, \quad \sigma\in[0,2\pi].\\
1 & \textrm{if $\sigma=2\pi$} \end{array}\right .
\end{equation}
Using standard Poisson algebra (\ref{eq:spa}) and transformation laws (\ref{eq:jed1})-(\ref{eq:jed3}), after one partial integration, we get
\begin{eqnarray}
&&\{\{y_\mu(\sigma_1),y_\nu(\sigma_2)\},y_\rho(\sigma_3)\}\nonumber\\
&&\quad=\frac{1}{2\kappa^2}H\varepsilon_{\mu\nu\rho}\left[\theta(\sigma_2-\sigma_1)\theta(\sigma_1-\sigma_3)\right.\nonumber\\
&&\quad+\left.\theta(\sigma_1-\sigma_2)\theta(\sigma_2-\sigma_3)\right]\, ,\label{eq:xy}
\end{eqnarray}

Now we have all ingredients to calculate the nonassociativity relation
\begin{eqnarray}
&&\{\{y_\mu(\sigma_1),y_\nu(\sigma_2)\},y_\rho(\sigma_3)\}\nonumber\\
&&\quad
\quad-\{y_\mu(\sigma_1),\{y_\nu(\sigma_2),y_\rho(\sigma_3)\}\}=\nonumber\\
&&\quad\frac{1}{2\kappa^2}H\varepsilon_{\mu\nu\rho}\left[2\theta(\sigma_3-\sigma_2)\theta(\sigma_2-\sigma_1)\right.\nonumber\\&&\quad+\left.\theta(\sigma_1-\sigma_3)\theta(\sigma_3-\sigma_2)\right.\nonumber\\&&\quad+\left.\theta(\sigma_3-\sigma_1)\theta(\sigma_1-\sigma_2)\right]
\end{eqnarray}
and breaking of Jacobi identity
\begin{eqnarray}
&&\{y_\mu(\sigma_1),y_\nu(\sigma_2),y_\rho(\sigma_3)\}\equiv\nonumber\\
&&\quad\{\{y_\mu(\sigma_1),y_\nu(\sigma_2)\},y_\rho(\sigma_3)\}\nonumber\\
&&\quad+\{\{y_\nu(\sigma_2),y_\rho(\sigma_3)\},y_\mu(\sigma_1)\}\nonumber\\
&&\quad+\{\{y_\rho(\sigma_3),y_\mu(\sigma_1)\},y_\nu(\sigma_2)\}=\nonumber \\
&&\quad \frac{1}{\kappa^2}H\varepsilon_{\mu\nu\rho}\left[\theta(\sigma_1-\sigma_2)\theta(\sigma_2-\sigma_3)\right.\nonumber\\&&\quad+\left.\theta(\sigma_3-\sigma_1)\theta(\sigma_1-\sigma_2)\right.\nonumber \\&&
\quad+ \left.\theta(\sigma_2-\sigma_3)\theta(\sigma_3-\sigma_1)\right]\, .
\end{eqnarray}
For example, for $\sigma_1=2\pi+\sigma$ and $\sigma_2=\sigma_3=\sigma$ one has
\begin{equation}
\{y_\mu(2\pi+\sigma),y_\nu(\sigma),y_\rho(\sigma)\}=-\frac{1}{\kappa^2}H\varepsilon_{\mu\nu\rho}\, .
\end{equation}

In approach of this article, the background of the T-dual theory depends on the non-local variable $V^\mu$ which incorporates main features of the non-geometric spaces.
Reducing our procedure to three dimensions and using the backgrounds of Refs. \cite{BDLPR,ALLP,RPEP}, we showed that our structure of arguments of background fields proves the proposal of Refs. \cite{BDLPR,RPEP} that
non-associativity and breaking of Jacobi identity are features of R-flux background.

\subsection{Critical surface}

Let us generalize the discussion of Ref. \cite{GRV} where the critical surface, which separates equivalent sections
of background fields, generalizes the critical radius.
Using the dilaton field analysis, namely the
relation  (\ref{eq:dr}),
we can conclude that
T-duality maps the theories with a given $$\det (2 \Pi_{\pm ab})$$  into the  theories with $${1 / \det (2 \Pi_{\pm ab})},$$
so that  all different theories are in the region $$\det (2 \Pi_{\pm ab}) \leq 0.$$ The theories which background fields satisfy the condition  $\det (2 \Pi_{\pm ab})=1$, are mapped into each other under T-duality.  This is a generalization of the critical radius and can be consider as a critical surface. So, relation (\ref{eq:dr}) implies $ \sqrt{\det G_{a b}} = \sqrt{\det {^\bullet G}^{a b}}$, which means that a dual volume 
is equal to the initial one. At the critical surface the extended symmetry  should be expected.

Let us, following \cite{GRV}, give an example of the relation between the original and T-dual background fields. We will consider the initial background  in 4-dimensional torus $T^4$ given by
\begin{equation}\label{eq:poc}
G_{\mu\nu}=g\delta_{\mu\nu},\quad
B_{\mu\nu}=b^{i}E^{i}_{\mu\nu},
\end{equation}
where 
\begin{eqnarray}
&&
E^{1}=
\begin{bmatrix}
\ 0&\ 0&\ 0 &\ 1\\
\ 0&\ 0&\ 1&\ 0\\
\ 0&-1&\ 0&\ 0\\
-1&\ 0&\ 0&\ 0
\end{bmatrix},
\nonumber\\
&&
E^{2}=\begin{bmatrix}
\ 0&\ 0&\ 1 &\ 0\\
\ 0&\ 0&\ 0&-1\\
-1&\ 0&\ 0&\ 0\\
\ 0&\ 1&\ 0&\ 0
\end{bmatrix},
\nonumber\\
&&
E^{3}=\begin{bmatrix}
\ 0&\ 1&\ 0 &\ 0\\
-1&\ 0&\ 0&\ 0\\
\ 0&\ 0&\ 0&\ 1\\
\ 0&\ 0&-1&\ 0
\end{bmatrix},\quad
\end{eqnarray}
satisfies
\begin{equation}
E^{i}E^{j}=-\delta^{ij}I+\varepsilon^{ijk}E^{k},\qquad \varepsilon^{123}=1.
\end{equation}
The zero modes  of the T-dual metric and T-dual Kalb-Ramond field  (\ref{eq:dpolja}) for the initial fields (\ref{eq:poc})
are
\begin{equation}
{^\star G}^{\mu\nu}=(G^{-1}_{E})^{\mu\nu}=\frac{g}{g^{2}+b^{2}}\,I,
\end{equation}
and
\begin{equation}
{^\star B}^{\mu\nu}=\frac{\kappa}{2}\theta^{\mu\nu}=
-\frac{1}{2}\frac{b^{i}}{g^{2}+b^{2}}\,E^{i},
\end{equation}
with $b^{2}=b^{i}b^{i}$.
They have the same form as the initial fields (\ref{eq:poc})
\begin{equation}
{^\star G}_{\mu\nu}={^\star g}\delta_{\mu\nu},\quad
{^\star B}_{\mu\nu}={^\star b}^{i}E^{i}_{\mu\nu},
\end{equation}
with
\begin{equation}\label{eq:dug}
{^\star g}=\frac{g}{g^{2}+b^{2}},
\quad
{^\star b}=-\frac{b^{i}}{g^{2}+b^{2}}.
\end{equation}
One easily shows
\begin{equation}
{^\star g}^{2}+{^\star b}^{2}=\frac{1}{g^{2}+b^{2}}.
\end{equation} 
In spheric coordinates one has
\begin{eqnarray}
&&(g,b^{1},b^{2},b^{3})=
(r\cos{\theta},r\sin{\theta}\cos{\varphi},
\nonumber\\
&&\quad r\sin{\theta}\sin{\varphi}\cos{\varphi_{1}},
r\sin{\theta}\sin{\varphi}\sin{\varphi_{1}}),
\end{eqnarray}
so $g^{2}+b^{2}=r^{2}$ and using (\ref{eq:dug}) one obtains
\begin{eqnarray}
&&({^\star g},
{^\star b}^{1},
{^\star b}^{2},
{^\star b}^{3})
=
(\frac{1}{r}\cos{\theta},
-\frac{1}{r}\sin{\theta}\cos{\varphi},
\nonumber\\
&&\quad
-\frac{1}{r}\sin{\theta}\sin{\varphi}\cos{\varphi_{1}},
-\frac{1}{r}\sin{\theta}\sin{\varphi}\sin{\varphi_{1}}).
\nonumber\\
\end{eqnarray}
Therefore, T-duality transforms $(r,\theta,\varphi,\varphi_{1})$ to
\begin{equation}
({^\star r},{^\star\theta},{^\star\varphi},{^\star\varphi_{1}})=(\frac{1}{r},-\theta,\varphi,\varphi_{1}).
\end{equation}

From the relation $\Pi_\pm G^{-1} \Pi_\mp = - \frac{1}{4} G_E$ we find 
\begin{equation}\label{eq:detpi}
\det (2 \Pi_{\pm \mu \nu})  = \frac{g^2}{{}^\star g^2} = (g^2 + b^2)^2 =r^4 \,  .
\end{equation}
Backgrounds corresponding to $r=1$ are mapped into themselves. The subset of this is the fixed surface with the condition
$$\det (2 \Pi_{\pm \mu \nu}) = r^4=1\, \, , \theta=0$$ or $g=1, \,\, b^{i}=0$.

\section{Conclusion}
\cleq

In this paper,
we considered the closed  string propagating  in the weakly curved background (\ref{eq:wcb}),
composed of a constant metric $G_{\mu\nu}$ and a linearly coordinate dependent Kalb-Ramond
field $B_{\mu\nu}$,
with an infinitesimal field strength.
We investigated the application of the generalized T-dualization procedure on the arbitrary set of coordinates and obtained the following T-duality diagram:

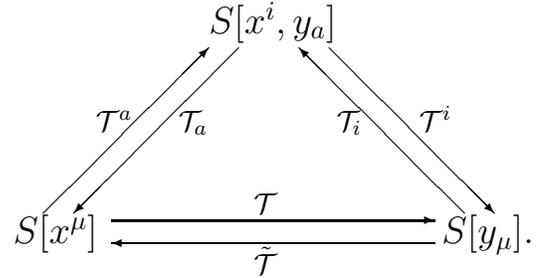
\begin{center}
\setlength{\unitlength}{1mm}
\begin{picture}(65,35)(5,5)
\put(1,1){{\Large$S[x^\mu]$}}
\put(14,4){\vector(1,0){43}}\put(58,1){{\Large$S[y_\mu]$.}}
\put(57,1){\vector(-1,0){43}}
\put(27,29){{\Large$S[x^i,y_a]$}}
\put(5,5){\vector(1,1){22}}
\put(31,27){\vector(-1,-1){22}}
\put(43,27){\vector(1,-1){22}}
\put(61,5){\vector(-1,1){22}}
\put(12,16){${\cal T}^a$}
\put(23,16){${\cal T}_a$}
\put(55,16){${\cal T}^i$}
\put(44,16){${\cal T}_i$}
\put(33,5){${\cal T}$}
\put(33,-3){${\tilde {\cal T}}$}
\end{picture}
\end{center}

\vskip 1cm

Let us stress that generalized T-dualization procedure enables the T-dualization along arbitrary direction,
even if the background fields depend on these directions.
The consequence of this procedure is that the arguments of the background fields,
such as $\Delta V^{a}$, are non-local.
They are non-local by definition, as they are the line integrals of the gauge fields.
Once the explicit form is obtained the non-locality is seen in a fact that they depend on double coordinates
$\tilde{x}$ and $\tilde{y}$, which are the line integrals of the $\tau$ and $\sigma$ derivatives of the original coordinates.
To all the theories considered,
except the initial theory, there corresponds the non-geometric, non-local flux.

The generalized T-dualization procedure
was first
applied along arbitrary  $d$ $(d=1,\dots,D-1)$ coordinates
$x^{a}=\{x^{\mu_{1}},\dots,x^{\mu_{d}}\}$.
We obtained the T-dual action $S[x^{i},y_{a}]$, given by eq. (\ref{eq:tdualdel})
with the dual background fields equal to
\begin{eqnarray}
&&{^\bullet}\Pi_{+ij}=\overline\Pi_{+ij},\quad
{^\bullet}\Pi_{+i}^{\ \ \ a}=-\kappa\Pi_{+ib}\tilde\Theta^{ba}_{-},
\nonumber\\
&&{^\bullet}\Pi^{a}_{+i}=\kappa\tilde\Theta^{ab}_{-}\Pi_{+bi},\quad
{^\bullet}\Pi_{+}^{ab}=\frac{\kappa}{2}\tilde\Theta^{ab}_{-}.
\end{eqnarray}
The argument of all background fields, $[x^i, V^a(x^i,y_a)]$,
depends nonlinearly on coordinates $x^i,y_a$ through their doubles $\tilde x^i,\tilde y_a$ [see (\ref{eq:vnula}) and (\ref{eq:tilds})]. 
All actions $S[x^i,y_a]$ are physically equivalent,
but are described with coordinates $x^{i}=\{x^{\mu_{d+1}},\dots,x^{\mu_{D}}\}$,
for the untreated directions 
and dual coordinates
$y_{a}=\{y_{\mu_{1}},\dots,y_{\mu_{d}}\}$, for the dualized directions.
The case $d=D$  corresponds to the completely T-dual action
with the T-dual fields $\frac{\kappa}{2}\Theta^{\mu\nu}_{-}\big(V(y)\big)$
and the case $d=0$ to the initial action with the background $\Pi_{+\mu\nu}(x)$.

Applying the procedure to the T-dual action along dual directions $y_{a}=\{y_{\mu_{1}},\dots,y_{\mu_{d}}\}$ we obtained the initial theory,
and applying it to the untreated directions $x^{i}=\{x^{\mu_{d+1}},\dots,x^{\mu_{D}}\}$
we obtained the completely T-dual theory.
All these derivations confirmed that
the set of all T-dualizations forms an Abelian group.
The neutral element of the group is the unexecuted T-dualization,
while the T-dualizations along some subset of original directions ${\cal T}^{a}$ is inverse
to the T-dualizations along the set of the corresponding dual directions ${\cal T}_{a}$.

\appendix

\section{The background field compositions}\label{sec:pp}
\cleq
The background field compositions $\Pi_{\pm \mu\nu}$ of the initial theory are
\begin{equation}
\Pi_{\pm \mu\nu}=B_{\mu\nu}\pm\frac{1}{2}G_{\mu\nu},
\end{equation}
where $G_{\mu\nu}$ and $B_{\mu\nu}$ are the initial metric and the initial Kalb-Ramond field.
The background field compositions ${\Theta}^{\mu\nu}_{\pm}$ of the T-dual theory are
\begin{eqnarray}\label{eq:tetaraz}
{\Theta}^{\mu\nu}_{\pm}&\equiv&
-\frac{2}{\kappa}
(G^{-1}_{E}\Pi_{\pm}G^{-1})^{\mu\nu}=
{\theta}^{\mu\nu}\mp \frac{1}{\kappa}(G_{E}^{-1})^{\mu\nu},
\nonumber\\
\end{eqnarray}
with $G_{E\mu\nu}$ being the effective metric
\begin{equation}
G_{E\mu\nu}\equiv G_{\mu\nu}-4(BG^{-1}B)_{\mu\nu},
\end{equation}
and $\theta^{\mu\nu}$ being the parameter of non-commutativity
\begin{equation}
\theta^{\mu\nu}\equiv
-\frac{2}{\kappa}
(G^{-1}_{E}BG^{-1})^{\mu\nu}.
\end{equation}
These background field compositions satisfy
\begin{equation}\label{eq:inv}
\Pi_{\pm \mu\nu}\Theta^{\nu\rho}_\mp=
\Theta^{\rho\nu}_\pm\Pi_{\mp\nu\mu}=\frac{1}{2\kappa}\delta^\rho_\mu.
\end{equation}

Let us define the analogs of ${\Theta}^{\mu\nu}_{\pm}$ in the $d$- and $D-d$-dimensional subspaces determined by coordinates 
$x^{a}=\{x^{\mu_{1}},\dots,x^{\mu_{d}}\}$ and
$x^{i}=\{x^{\mu_{d+1}},\dots,x^{\mu_D}\}$, where
$d=1,2,\dots,D-1$.
The effective metrics in these subspaces are defined by 
\begin{eqnarray}\label{eq:effmet}
&&\tilde{G}_{Eab}\equiv G_{ab}-4B_{ac}(\tilde{G}^{-1})^{cd}B_{db},
\nonumber\\
&&\bar{G}_{Eij}\equiv G_{ij}-4B_{ik}(\bar{G}^{-1})^{kl}B_{lj},
\end{eqnarray}
where $\tilde{G}_{ab}\equiv G_{ab}$ and $\bar{G}_{ij}\equiv G_{ij}$.
Using these we define the
following field compositions:
\begin{eqnarray}\label{eq:barteta1}
&&\tilde{\Theta}^{ab}_{\pm}\equiv-\frac{2}{\kappa}
(\tilde{G}_{E}^{-1})^{ac}\Pi_{\pm cd}(\tilde{G}^{-1})^{db},
\nonumber\\
 &&\bar{\Theta}^{ij}_{\pm}\equiv-\frac{2}{\kappa}
(\bar{G}_{E}^{-1})^{ik}\Pi_{\pm kl}(\bar{G}^{-1})^{lj},
\end{eqnarray}
which are in fact the inverses of $2\kappa\Pi_{\mp ab}$ and $2\kappa\Pi_{\mp ij}$
\begin{eqnarray}\label{eq:tbarp1}
&&\tilde\Theta^{ab}_\pm\Pi_{\mp bc}=
\Pi_{\mp cb}\tilde\Theta^{ba}_\pm=
\frac{1}{2\kappa}\delta^{a}_{c},
\nonumber\\
&&\bar\Theta^{ij}_\pm\Pi_{\mp jk}=
\Pi_{\mp kj}\bar\Theta^{ji}_\pm=
\frac{1}{2\kappa}\delta^{i}_{k}.
\end{eqnarray} 
Analogously as the fields theta $\Theta^{\mu\nu}_\pm$ defined in the whole space by (\ref{eq:tetaraz}),
the theta fields defined in the subspaces
can be separated into antisymmetric and symmetric parts as
\begin{eqnarray}
&&\tilde{\Theta}^{ab}_{\pm}=\tilde\theta^{ab}\mp\frac{1}{\kappa}(\tilde{G}_{E}^{-1})^{ab},
\quad
\nonumber\\
&&\bar{\Theta}^{ij}_{\pm}=\bar\theta^{ij}\mp\frac{1}{\kappa}(\bar{G}_{E}^{-1})^{ij},
\end{eqnarray}
where
\begin{eqnarray}\label{eq:tetamalo}
&&\tilde\theta^{ab}\equiv
-\frac{2}{\kappa}
(\tilde{G}_{E}^{-1})^{ac}B_{cd}(\tilde{G}^{-1})^{db},
\nonumber\\
&&
\bar\theta^{ij}\equiv
-\frac{2}{\kappa}
(\bar{G}_{E}^{-1})^{ik}B_{kl}(\bar{G}^{-1})^{lj}.
\end{eqnarray}

In the zeroth order the quantities $\Pi_{\pm\mu\nu}$, $\Theta^{\mu\nu}_\pm$, $\widetilde\Theta^{ab}_\pm$ and $\bar\Theta^{ij}_\pm$ reduce to
\begin{eqnarray}\label{eq:nulte}
&&\Pi_{0\pm\mu\nu}=b_{\mu\nu}\pm\frac{1}{2}G_{\mu\nu},
\nonumber\\
&&\Theta^{\mu\nu}_{0\pm}=
-\frac{2}{\kappa}(g^{-1})^{\mu\rho}\Pi_{0\pm\rho\sigma}(G^{-1})^{\sigma\nu}=
{\theta}^{\mu\nu}_{0}\mp \frac{1}{\kappa}(g^{-1})^{\mu\nu},
\nonumber\\
&&\widetilde\Theta^{ab}_{0\pm}=
-\frac{2}{\kappa}({\tilde{g}}^{-1})^{ac}\,\Pi_{0\pm cd}({\tilde{G}}^{-1})^{db}
=\tilde{\theta}^{ab}_{0}\mp \frac{1}{\kappa}({\tilde{g}}^{-1})^{ab},
\nonumber\\
&&\bar\Theta^{ij}_{0\pm}=
-\frac{2}{\kappa}({\bar{g}}^{-1})^{ik}\,\Pi_{0\pm kl}({\bar{G}}^{-1})^{lj}
=\bar{\theta}^{ij}_{0}\mp \frac{1}{\kappa}({\bar{g}}^{-1})^{ij},
\nonumber\\
\end{eqnarray}
where the zeroth order effective metrics are
\begin{eqnarray}
&&
g_{\mu\nu}=G_{\mu\nu}-4b_{\mu\rho}(G^{-1})^{\rho\sigma}b_{\sigma\nu},
\nonumber\\
&&{\tilde{g}}_{ab}=G_{ab}-4b_{ac}(\tilde{G}^{-1})^{cd}b_{db},
\nonumber\\
&&{\bar{g}}_{ij}=G_{ij}-4b_{ik}(\bar{G}^{-1})^{kl}b_{lj},
\end{eqnarray}
and
the zeroth order non-commutativity parameters are
\begin{eqnarray}
&& {\theta}^{\mu\nu}_{0}=
-\frac{2}{\kappa}(g^{-1})^{\mu\rho}b_{\rho\sigma}(G^{-1})^{\sigma\nu},
\nonumber\\
&& \tilde{\theta}^{ab}_{0}=
-\frac{2}{\kappa}({\tilde{g}}^{-1})^{ac}\,b_{cd}({\tilde{G}}^{-1})^{db}
\nonumber\\
&& \bar{\theta}^{ij}_{0}=
-\frac{2}{\kappa}({\bar{g}}^{-1})^{ik}\,b_{kl}({\bar{G}}^{-1})^{lj}.
\end{eqnarray}
Quantities $\Pi_{0\pm\mu\nu}$, $\Theta^{\mu\nu}_{0\pm}$, $\widetilde\Theta^{ab}_{0\pm}$ and $\bar\Theta^{ij}_{0\pm}$ satisfy
\begin{eqnarray}\label{eq:ptnula}
&&\Pi_{0\pm \mu\nu}\Theta^{\nu\rho}_{0\mp}=
\Theta^{\rho\nu}_{0\pm}\Pi_{0\mp\nu\mu}=\frac{1}{2\kappa}\delta^\rho_\mu,
\nonumber\\
&&
\Pi_{0\pm ab}\tilde\Theta^{bc}_{0\mp}=
\tilde\Theta^{cb}_{0\pm}\Pi_{0\mp ba}=
\frac{1}{2\kappa}\delta^{c}_{a},
\nonumber\\
&&
\Pi_{0\pm ij}\bar\Theta^{jk}_{0\mp}=
\bar\Theta^{kj}_{0\pm}\Pi_{0\mp ji}=
\frac{1}{2\kappa}\delta^{k}_{i}.
\end{eqnarray}

The non-commutativity parameters
theta $\Theta^{\mu\nu}_\pm$, $\tilde\Theta^{ab}_\pm$ and $\bar\Theta^{ij}_\pm$ can be expressed as
\begin{eqnarray}\label{eq:tetarazvoj}
&&\Theta^{\mu\nu}_\pm=
\Theta^{\mu\nu}_{0\pm}
-2\kappa\Theta^{\mu\rho}_{0\pm}h_{\rho\sigma}\Theta^{\sigma\nu}_{0\pm},
\nonumber\\
&&\tilde{\Theta}^{ab}_{\pm}=\tilde{\Theta}^{ab}_{0\pm}
-2\kappa\tilde{\Theta}^{ac}_{0\pm}h_{cd}\tilde{\Theta}^{db}_{0\pm},
\nonumber\\
&&\bar{\Theta}^{ij}_{\pm}=\bar{\Theta}^{ij}_{0\pm}
-2\kappa\bar{\Theta}^{ik}_{0\pm}h_{kl}\bar{\Theta}^{lj}_{0\pm}.
\end{eqnarray}

\subsection{Relations between field compositions}

In Sect. \ref{sec:pscrtom} we introduced the background field composition
\begin{equation}\label{eq:pcrta}
\overline{\Pi}_{\pm ij}\equiv
\Pi_{\pm ij}
-2\kappa\Pi_{\pm ia}\tilde\Theta^{ab}_{\mp}\Pi_{\pm bj},
\end{equation}
and analogously we define
\begin{equation}\label{eq:ptilda}
\widetilde{{\Pi}}_{\pm ab}\equiv
\Pi_{\pm ab}
-2\kappa\Pi_{\pm ai}\bar\Theta^{ij}_{\mp}\Pi_{\pm jb}.
\end{equation}
Here we will show that these quantities are the inverses of the ordinary
non-commutativity parameters  theta,
projected to $i$ and $a$-subspaces [see (\ref{eq:relacije})].

Let us
express tensors $\Pi_{\pm \mu\nu}$ and $\Theta^{\mu\nu}_\pm$,
which satisfy (\ref{eq:inv})
in a block-wise form as 
\begin{equation}\label{eq:ptbw}
\Pi_{\pm \mu\nu}=\left(
\begin{array}{cc}
\Pi_{\pm ij} & \Pi_{\pm ib}\\
\Pi_{\pm aj} & \Pi_{\pm ab}
\end{array}\right)\, ,\quad \Theta^{\mu\nu}_\pm=
\left(
\begin{array}{cc}
\Theta_{\pm}^{ij} & \Theta_\pm^{ib}\\
\Theta_\pm^{aj} & \Theta_\pm^{ab}
\end{array}\right)\,.
\end{equation}
We will use the  
definition of block-wise inversion which states that the inverse of the matrix of the form
\begin{equation}
M=\left(\begin{array}{cc}
A & B\\
C & D         
\end{array}\right)\,
\end{equation}
equals
\begin{eqnarray}\label{eq:inverz}
&&M^{-1}=
\nonumber\\\nonumber\\
&&\!\!\!\!\!\!\!\!\left(
\begin{array}{cc}
(A-BD^{-1}C)^{-1} & -A^{-1}B(D-CA^{-1}B)^{-1}\\
-D^{-1}C(A-BD^{-1}C)^{-1} & (D-CA^{-1}B)^{-1}
\end{array}\right).
\nonumber\\
\end{eqnarray}
Applying (\ref{eq:inverz}) to the first matrix in (\ref{eq:ptbw}),
the relation (\ref{eq:inv}) implies
\begin{eqnarray}
&&2\kappa\Theta_{\mp}^{ij}=
\big(\Pi_{\pm ij}-2\kappa\Pi_{\pm ia}\tilde\Theta_\mp^{ab}\Pi_{\pm bj}\big)^{-1},
\nonumber\\
&&2\kappa\Theta_\mp^{ib}=-2\kappa\bar\Theta^{ij}_{\mp}
\Pi_{\pm ja}(\Pi_{\pm ab}-2\kappa\Pi_{\pm ak}
\bar\Theta^{kl}_{\mp}
\Pi_{\pm lb})^{-1},
\nonumber\\
&&2\kappa\Theta_\mp^{aj}=-2\kappa\tilde\Theta_\mp^{ab}\Pi_{\pm bi}(\Pi_{\pm ij}-2\kappa\Pi_{\pm ic}\tilde\Theta_\mp^{cd}\Pi_{\pm dj})^{-1},
\nonumber\\
&&2\kappa\Theta_\mp^{ab}=(\Pi_{\pm ab}-2\kappa\Pi_{\pm ai}
\bar\Theta^{ij}_{\mp}
\Pi_{\pm jb})^{-1},
\end{eqnarray}
and we can conclude that (\ref{eq:pcrta}) and (\ref{eq:ptilda}) are inverses of $2\kappa\Theta_{\mp}^{ij}$ and $2\kappa\Theta_{\mp}^{ab}$ respectively.
So, we can write
\begin{eqnarray}\label{eq:relacije}
&&{\overline{\Pi}}_{\pm ij}\Theta^{jk}_\mp=
\Theta^{kj}_\mp{\overline{\Pi}}_{\pm ji}=\frac{1}{2\kappa}\delta^{k}_{i},
\nonumber\\
&&
{\widetilde{\Pi}}_{\pm ab}\Theta^{bc}_\mp=
\Theta^{cb}_\mp{\widetilde{\Pi}}_{\pm ba}=\frac{1}{2\kappa}\delta^{c}_{a},
\end{eqnarray}
and
\begin{eqnarray}\label{eq:tetaia}
&&\Theta_{\mp}^{ib}=
-2\kappa\bar\Theta^{ij}_{\mp}\Pi_{\pm ja}\Theta_{\mp}^{ ab},
\nonumber\\
&&\Theta_{\mp}^{aj}=
-2\kappa\tilde\Theta_\mp^{ab}\Pi_{\pm bi}\Theta_{\mp}^{ij}.
\end{eqnarray}
Applying (\ref{eq:inverz}) to the second matrix in (\ref{eq:ptbw}),
Eq. (\ref{eq:inv}) implies
\begin{eqnarray}
&&2\kappa\Pi_{\mp ij}=(\Theta_{\pm}^{ij}
-2\kappa\Theta_\pm^{ia}
\widetilde\Pi_{ab\mp}
\Theta_\pm^{bj})^{-1},
\nonumber\\
&&2\kappa\Pi_{\mp ib}=-2\kappa\overline\Pi_{\mp ij}
\Theta_\pm^{ja}(\Theta_\pm^{ab}-2\kappa\Theta_\pm^{ak}
\overline\Pi_{\mp kl}
\Theta_\pm^{lb})^{-1},
\nonumber\\
&&2\kappa\Pi_{\mp aj}=-2\kappa
\widetilde\Pi_{ab\mp}
\Theta_\pm^{bi}(\Theta_{\pm}^{ij}-2\kappa\Theta_\pm^{ic}
\widetilde\Pi_{\mp cd}
\Theta_\pm^{dj})^{-1},
\nonumber\\
&&2\kappa\Pi_{\mp ab}=(\Theta_\pm^{ab}-2\kappa\Theta_\pm^{ai}
\overline\Pi_{\mp ij}
\Theta_\pm^{jb})^{-1},
\end{eqnarray}
so using (\ref{eq:tbarp1}) we conclude that
\begin{eqnarray}\label{eq:bttt}
&&\bar\Theta_{\pm}^{ij}=
\Theta_{\pm}^{ij}
-2\kappa\Theta_\pm^{ia}
\widetilde\Pi_{ab\mp}
\Theta_\pm^{bj},
\nonumber\\
&&
\tilde\Theta_\pm^{ab}=
\Theta_\pm^{ab}-2\kappa\Theta_\pm^{ai}
\overline\Pi_{\mp ij}\Theta_\pm^{jb},
\end{eqnarray}
and
\begin{eqnarray}\label{eq:pidiag}
&&\Pi_{\mp ib}=-2\kappa
\overline\Pi_{\mp ij}
\Theta_{\pm}^{ja}\Pi_{\mp ab},
\nonumber\\
&&
\Pi_{\mp aj}=-2\kappa
\widetilde\Pi_{\mp ab}
\Theta_\pm^{bi}\Pi_{\mp ij}.
\end{eqnarray}

Let us derive some useful relations between these quantities.
Equation (\ref{eq:inv}),
for $\mu=a,\,\nu=i$ and $\mu=i,\,\nu=a$ becomes
\begin{eqnarray}\label{eq:rel1}
&&\Pi_{\pm ab}\Theta_\mp^{bi}=-\Pi_{\pm aj}\Theta_\mp^{ji},
\nonumber\\
&&\Pi_{\pm ij}\Theta_\mp^{ja}=-\Pi_{\pm ib}\Theta_\mp^{ba},
\end{eqnarray}
while taking $\mu=a,\,\nu=b$ and $\mu=i,\,\nu=j$ we obtain
\begin{eqnarray}\label{eq:pomoc}
&&\Pi_{\pm ac}\Theta^{cb}_\mp+\Pi_{\pm ai}\Theta^{ib}_\mp=
\frac{1}{2\kappa}\delta^{b}_{a},
\nonumber\\
&&
\Pi_{\pm ia}\Theta^{aj}_\mp+
\Pi_{\pm ik}\Theta^{kj}_\mp
=\frac{1}{2\kappa}\delta^{j}_{i}.
\end{eqnarray}

Multiplying Eq. (\ref{eq:rel1}) from the left with $\tilde\Theta_{\mp}^{ca}$ and from the right with $\bar\Pi_{\mp ik}$ we get the relation
\begin{equation}\label{eq:tbpttp}
\Theta^{ci}_\mp \overline\Pi_{\pm ik}=-
\tilde\Theta^{ca}_\mp \Pi_{\pm ak}\, ,
\end{equation}
while
multiplying Eq.
(\ref{eq:pomoc})
 from the right with $\bar\Theta_{\mp}^{ki}$ and from the left with $\tilde\Pi_{\pm ac}$, we obtain
\begin{equation}
\Theta^{ka}_\mp \widetilde\Pi_{\pm ac}=-
\bar\Theta^{ki}_{\mp}\Pi_{\pm ic}.
\end{equation}


\end{document}